\documentclass[preprint, preprintnumbers, 12pt, onecolumn, amsfonts, aps, prc, nofootinbib, floatfix]{revtex4-1}

\usepackage{amsmath}
\usepackage{mathtools}
\usepackage{wasysym}

\def\l{\left}
\def\r{\right}

\def\ch{\mathrm{ch}}
\def\pp{$pp$\,}
\def\pPb{$p$-Pb\,}
\def\PbPb{Pb-Pb\,}
\def\pA{$p$-$A$\,}
\def\AA{$A$-$A$\,}

\begin{document}
\count\footins = 1000
\title{Hanbury Brown--Twiss Interferometry and Collectivity in Small Systems}

\author{Christopher Plumberg}
\affiliation{Theoretical Particle Physics,
Department of Astronomy and Theoretical Physics,\\
Lund University, S\"olvegatan 14A, SE-223 62 Lund, Sweden}

\preprint{LU TP 20-44}

\begin{abstract}
Hanbury Brown--Twiss interferometry (HBT) provides crucial insights into both the space-time structure and the momentum-space evolution of ultrarelativistic nuclear collisions at freeze-out.  In particular, the dependence of the HBT radii on the transverse pair momentum $K_T$ and the system charged multiplicity $dN_\ch/d\eta$ may reflect the mechanisms driving collective behavior in small systems.  This paper argues that certain features observed in the multiplicity dependence of the HBT radii can be naturally understood if small systems evolve hydrodynamically at high-multiplicity.  This study thus establishes a baseline for the multiplicity dependence of HBT in hydrodynamics which may prove useful in discriminating between competing models of collectivity in nuclear collisions.

\end{abstract}

\date{\today}

\maketitle

\section{Introduction}
\label{Sec1}

The existence of collective, fluid-like behavior in relativistic nuclear collisions, from \pp to \AA, is by now well established \cite{Nagle:2018nvi, Adolfsson:2020dhm}.  Understanding the precise origins of this collective behavior, however, remains one of the foremost outstanding challenges in the field.  To date, a number of explanations of this phenomenon have been proposed, including CGC-type models with collectivity built into the initial state \cite{Martinez:2018tuf, Wertepny:2020jun}, ``escape mechanism" models which effectively generate collectivity kinematically \cite{Lin:2015ucn}, approaches based on string hadronization models \cite{Bierlich:2017vhg, Sjostrand:2018xcd}, ``one-hit" dynamical models \cite{Kurkela:2018ygx}, and relativistic hydrodynamics \cite{Schenke:2010nt, *Gale:2012rq, *Gale:2013da, Shen:2014vra, Weller:2017tsr}.\footnote{It is, of course, possible to have combinations of these or other more basic approaches as well; see \cite{Schenke:2019pmk} for a recent example.}

The ability to discriminate between competing models of collectivity is therefore urgently needed and requires both quantitative predictions and comparison with experiment.  In this context, femtoscopic observables, such as those derived from Hanbury Brown--Twiss (HBT) interferometry, offer a powerful and complementary glimpse into the space-time structure and dynamical evolution of nuclear collisions at freeze-out \cite{Lisa:2005dd}.  The most widely used of these observables, the ``HBT radii," reflect collective effects in a number of ways, particularly in their dependence on the transverse pair momentum and on the system's charged multiplicity \cite{Makhlin:1987gm, Hirono:2014dda, Graef:2012sh, *Li:2012np, Heinz:2019dbd}.  For instance, when comparing large and small systems at fixed multiplicity, initially more compact systems (\pp, \pA) need to develop stronger transverse flow in order to reach the same final freeze-out volume as attained in larger systems \cite{Heinz:2019dbd}.  This enhanced flow, which is driven by the larger initial density gradients present in small systems \cite{Kalaydzhyan:2015xba}, is a direct prediction of hydrodynamics and leads to a measurable ordering \pp $<$ \pA $<$ \AA in the radii extracted from different systems at the same multiplicity.  This study will explore the implications of the enhanced radial flow produced by hydrodynamics in small systems for the multiplicity dependence of the HBT radii.

The multiplicity dependence of HBT in small and large systems has already been studied in a fair number of experimental analyses \cite{Kisiel:2011jg, Aamodt:2011kd, Adamczyk:2014mxp, Adam:2015pya}.  One notable feature of these measurements is that the collision system's volume, when estimated from the HBT radii, appears to scale linearly with the charged particle multiplicity $dN_\ch/d\eta$, while the individual radii each scale linearly with $(dN_\ch/d\eta)^{1/3}$.  Moreover, the dependence of the individual radii in \textit{large} systems is seen to fall roughly onto a single universal, approximately linear trajectory in $(dN_\ch/d\eta)^{1/3}$, regardless of collision species \cite{Lisa:2005dd}.  This is exactly what one would expect to find if \pp, \pA, and \AA collisions are all driven by hydrodynamics.

What is initially surprising, then, is to find in the data that the individual radii across collision systems should differ not only in their \textit{magnitudes}, as implied by the enhanced flow in small systems, but also in the \textit{slopes} of their respective $(dN_\ch/d\eta)^{1/3}$ dependences.  There are two specific features to be noted.  First, in large systems, as noted above, each radius has a slope which is approximately independent of the collision species, whereas in \pp and \pA, the corresponding slopes tend to be considerably smaller (the exception is the `side' radius, as we will see below).  I will refer to this feature of the data as the \textit{slope non-universality} exhibited by the radii in small systems.  In addition, not only do the slopes in small systems tend to deviate from the universal slope of large systems, but they also disagree more significantly amongst themselves: the rough ordering of the slopes within a fixed system is
\begin{eqnarray*}
	A-A &:& \text{out}>\text{long}\sim\text{side} \\
	p-A &:& \text{long}\gtrsim\text{side}\sim\text{out}\\
	pp &:& \text{side}>\text{long}\sim\text{out}
\end{eqnarray*}
There is therefore also a \textit{slope hierarchy} exhibited by the different radii which depends on which collision system is being considered.  This hierarchy is clearly present in both large and small systems, but varies in strength between them.  The presence of these features in the multiplicity dependence of HBT in \pp, \pA, and \AA implies radical differences in the space-time evolution of large and small systems.

The preceding observations seem to introduce some unwelcome complexity into an otherwise simple situation.  Taken together, the two features just identified -- the non-universality and hierarchy of the slopes -- appear to stand in tension with the intuitive expectation that hydrodynamics should lead systems of all sizes to evolve in similar ways.  This in turn raises the crucial question of whether the features appearing in small systems are indeed signatures of genuine, hydrodynamic collectivity or of something else and, more generally, whether the mechanisms driving the dynamical evolution of small systems are the same as those at play in larger systems \cite{Bozek:2013df, McLerran:2013xba}.

The goal of this paper is to show how the features just identified in the multiplicity dependence of the HBT radii arises naturally within the context of hydrodynamics.  More precisely, I will show using a simplified hydrodynamic model that both the non-universality and the hierarchy exhibited by the slopes of the $(dN_\ch/d\eta)^{1/3}$ dependence of the HBT radii across different collision systems emerges naturally within the hydrodynamic paradigm, at least at sufficiently large multiplicities.  While the simplifications used in my hydrodynamic modeling limit my discussion here to a somewhat qualitative level, they can (and should) be removed for a more quantitative interpretation of the experimental data in future work.

The outline of this paper is as follows.  In Sec.~\ref{Sec2} I review the basic elements of HBT interferometry and show how to justify the interpretation of the HBT radii in terms of the space-time structure of the freeze-out surface.  In Sec.~\ref{Sec3}, I take a closer look at the data which most clearly illustrate the significant differences in the radii and their multiplicity scaling when compared across various collision systems.  In Sec.~\ref{Sec4} I will show using a highly simplified hydrodynamic model how the discrepancies observed in Sec.~\ref{Sec3} might reflect fluid dynamical behavior in small collision systems.  Finally, Sec.~\ref{Sec5} will summarize the main results and offer some suggestions for future work which will flesh out these ideas in a more quantitative fashion.

\section{Formalism}
\label{Sec2}

\subsection{The correlation function}
HBT interferometry relies on the existence of Bose-Einstein or Fermi-Dirac correlations between pairs of identical particles.  The techniques underlying HBT have been developed and reviewed extensively elsewhere \cite{Wiedemann:1999qn, Lisa:2005dd}.  Here, I will briefly present the essential elements which are necessary to establish my notation and to show how the space-time structure of the source may be inferred.

The basic observable of HBT interferometry is the two-particle correlation function, defined by
\begin{equation}
	C(\vec{p}_1, \vec{p}_2)
	= \frac{E_1 E_2 \frac{dN}{d^3p_1 d^3p_2}}
	       {\l(E_1 \frac{dN}{d^3p_1}\r)
	        \l(E_2 \frac{dN}{d^3p_2}\r)}.
	\label{2pCF_def}
\end{equation}
Ideally, it is constructed so as to reduce to unity in the absence of actual Bose-Einstein correlations between identical particle pairs -- in our case, pairs of $\pi^+$ bosons -- produced by the collision event.  Theoretically, it is usually convenient to consider instead of \eqref{2pCF_def} the equivalent correlation function evaluated in terms of the relative momentum $q = p_1 - p_2$ and the pair momentum $K = (p_1 + p_2)/2$ and relate it directly to an ``emission function" (or ``source function") $S(x,K)$:
\begin{eqnarray}
 C(\vec{q}, \vec{K})
  &=& 1 + \frac{\l| \int d^4x e^{i q \cdot x} S(x,K) \r|^2}{\l( \int d^4x S(x,K+\frac{q}{2}) \r)\l( \int d^4y S(y,K-\frac{q}{2}) \r)}
  \label{C_q_K_vs_S_x_K_definition_woSA}\\
  &\approx & 1 + \frac{\l| \int d^4x e^{i q \cdot x} S(x,K) \r|^2}{\l| \int d^4x S(x,K) \r|^2}
  \label{C_q_K_vs_S_x_K_definition_wSA}
\end{eqnarray}
$S$ can be thought of in essence as a quantum-mechanical phase space (or `Wigner') distribution \cite{Heinz:1996bs} which roughly characterizes the probability to emit a particle from position $x$ with momentum $K$.  The step from \eqref{C_q_K_vs_S_x_K_definition_woSA} to \eqref{C_q_K_vs_S_x_K_definition_wSA} makes use of the so-called ``smoothness assumption'' \cite{Pratt:1997pw} which is well-justified for large sources such as \AA collisions, but becomes questionable in \pp and \pA collisions.  This assumption will be relaxed in the full analysis which follows, although its effects turn out to be mostly negligible when evaluated quantitatively.

In any event, the width of the correlation function in $\vec{q}$ reflects the space-time structure of the underlying source at a fixed $\vec{K}$.  This structure can be inferred by suitably parameterizing the correlation function with a functional form such as
\begin{equation}
  C_{fit}(\vec{q}, \vec{K})
   = 1 + \lambda(\vec{K}) \exp\l( - \sum_{i,j \in \l\lbrace o,s,l \r\rbrace} R^2_{ij}(\vec{K}) q_i q_j \r).
   \label{fitCF}
\end{equation}
The $R^2_{ij}(\vec{K})$ and $\lambda(\vec{K})$ are extracted as free parameters, obtained by fitting \eqref{fitCF} to one of the theoretical correlation functions \eqref{C_q_K_vs_S_x_K_definition_woSA} or \eqref{C_q_K_vs_S_x_K_definition_wSA}.  The quantities $R^2_{ij}(\vec{K})$ are known as the HBT radii and quantify the space-time structure of the emitting source, and $\lambda(\vec{K})$ is an \textit{ad hoc} factor which typically deviates from unity when effects due to resonance decays \cite{Wiedemann:1996ig} or coherent pion production \cite{Sinyukov:1994en, Sinyukov:2012ut, Akkelin:2011zz, Shapoval:2013jca} are important.  These effects will be neglected in this study, meaning that $\lambda(\vec{K})=1$.\footnote{Eq.~\eqref{fitCF} also neglects the effects of Coulomb and other final-state interactions, which are assumed to be corrected for at the level of the experimental analysis.}

In the special case of a perfectly Gaussian source (and making use of the smoothness assumption) \cite{Heinz:1996bs}, one can perform the Fourier integrals in \eqref{C_q_K_vs_S_x_K_definition_wSA} analytically, yielding an exact relation between the $R^2_{ij}(\vec{K})$ and space-time variances of the underlying source \cite{Heinz:1999rw, Heinz:2004qz}:
\begin{eqnarray}
	R^2_{ij}(\vec{K}) &=& \l< (\tilde{x}_i - \beta_i \tilde{t})(\tilde{x}_j - \beta_j \tilde{t}) \r>, \label{GSA_rel1}\\
	\tilde{x}_i &=& x_i - \l<x_i\r>, \tilde{t} = t - \l<t\r>. \label{GSA_rel2}
\end{eqnarray}
Here the averages are taken with respect to the emission function:
\begin{equation}
	\l< g(x) \r> \equiv \frac{\int d^4x\, g(x) S(x,K)}{\int d^4x\, S(x,K)}. \label{GSA_rel3}
\end{equation}
Additionally, the pair velocity $\vec{\beta}$ is given by
\begin{equation}
\vec{\beta} = \frac{\vec{K}}{K^0}
	\approx \frac{\vec{K}}{\sqrt{m_\pi^2 + \vec{K}^2}}, \label{beta_def}
\end{equation}
and the separate terms ($\l< \tilde{x}_i \tilde{x}_j \r>$, $\ldots$) comprising the righthand side of Eq.~\eqref{GSA_rel1} are known as the ``source variances" \cite{Plumberg:2015eia}.  Although the ``Gaussian source approximation" is not used in this analysis, it will be useful here in interpreting and developing intuition for the results presented below.

It is worth emphasizing that the relations \eqref{GSA_rel1} - \eqref{GSA_rel3} provide the essential connection between the $R^2_{ij}$ and the spatial and temporal characteristics of the underlying source function, which justifies the usual interpretation of the HBT radii in terms of the space-time geometry of the collision system at freeze-out.  Nevertheless, the radii do not reflect \textit{only} spatial lengthscales in the system, but necessarily represent a mixture of spatial and temporal information together.

\subsection{The emission function}
To proceed further, we need to specify the emission function $S$ which governs the particle production process in a nuclear collision.  For the systems studied here using hydrodynamics, the emission function can be defined straightforwardly according to the standard Cooper-Frye prescription \cite{Cooper:1974mv}:
\begin{eqnarray*}
	S(x,p) &=& \frac{1}{(2\pi)^3} \int_{\Sigma(y)}
	           p \cdot d^3 \sigma(y) f(y, p)
	                 \delta^4 (x-y)\\
	f(x,p) &=& \frac{1}
	           {e^{(p \cdot u(x) - \mu)/T} \pm 1},
	\label{CooperFrye}
\end{eqnarray*}
where $u(x)$ is the local flow velocity profile and $\Sigma(y)$ signifies the freeze-out surface over which the integral is evaluated.  Viscous corrections to the distribution function have been neglected here for simplicity.  This is reasonable, as the precise form and magnitude of these corrections are still not extremely well-constrained theoretically \cite{McNelis:2019auj, Paquet:2020rxl}, and in any event have little effect on the qualitative behavior of the $R^2_{ij}$ obtained from hydrodynamic simulations with smooth or event-averaged initial states \cite{Plumberg:2015eia}.\footnote{The same is not necessarily true for the hydrodynamic simulations themselves, as viscous effects influence not only the amount of particle production (and therefore the final charged multiplicity) but also contribute to the transverse flow \cite{Teaney:2003kp} and consequently affect the shape of the freeze-out surface as well.  For these reasons, some viscous effects have been retained in the hydrodynamic simulations presented below, despite being excluded from the calculation of the HBT radii.}

\subsection{Initial Conditions and Hydrodynamics}
Hydrodynamics of course requires initial conditions.  The initial conditions for this analysis were generated using the MC-Glauber  model \cite{Loizides:2014vua, *Bozek:2019wyr}, including fluctuations of both the nucleon positions and collision-by-collision multiplicity fluctuations \cite{Shen:2014vra}.  Event-averaged initial conditions were generated for each system (\pp, \pPb, \PbPb) in 10\% centrality-class intervals ($0-10\%, \ldots 90-100\%$) by cutting on the total initial entropy at mid-rapidity, as described also in \cite{Shen:2014vra}.  The overall normalization for each system was adjusted so that the system yielded a benchmark value of the charged particle pseudorapidity density $dN_\ch/d\eta$, obtained from experimental measurements for that system in a given reference centrality class.  The benchmark value of $dN_\ch/d\eta$ in each system's respective reference class is given in Table \ref{Table1}.  Several higher centrality classes (0-1\%, 0-0.1\%, 0-0.01\%, and 0-0.001\%) were also generated for \pp and \pPb.
\begin{table}[]
\begin{tabular}{|l|r|c|c|}
\hline
\multicolumn{1}{|c|}{System} & \multicolumn{1}{|c|}{$\sqrt{s_{NN}}$} & \,Reference Class\, & $dN_\ch/d\eta$ \\ \hline
\,\pp & 7 TeV\, & 0-100\% & 6.0 \\ \hline
\,\pPb & 5.02 TeV\, & 0-100\% & 17.5 \\ \hline
\,\PbPb \, & \,2.76 TeV\, & 0-5\% & 1601 \\ \hline
\end{tabular}
\caption{The target multiplicities for each system considered in the respective centrality classes shown. 
The target values are chosen to agree approximately with the measurements presented in Refs.~\cite{Adam:2015gka} (\pp), \cite{ALICE:2012xs} (\pPb), and \cite{Aamodt:2010cz, *Abbas:2013bpa} (\PbPb).  
\label{Table1}}
\end{table}

The hydrodynamic evolution was performed using the 2+1D iEBE-VISHNU package \cite{Shen:2014vra} with specific viscosities $\eta/s = 0.08$ and $\zeta/s = 0$ and the \textit{s95p}-v1 equation of state \cite{Huovinen:2009yb}.  Hydrodynamics was initialized at a proper time $\tau_0 = 0.6$ fm/$c$ without including any preequilibrium effects.  This is already a significant assumption, especially for small systems: hydrodynamics is typically valid only after preequilibrium dynamics (e.g., \cite{Kurkela:2018vqr}) have enabled the system to `hydrodynamize' on a timescale $\tau_{\mathrm{hydro}} \sim O(1/T)$ set by the temperature, which varies with the size of the collision system \cite{Busza:2018rrf, *Berges:2020fwq}.  However, the fact that this simplification omits an important source of transverse flow in nuclear collisions means it is likely to \textit{underestimate} the effects on the HBT radii \cite{Pratt:2008qv}.  Since including preequilibrium flow would likely only strengthen the conclusions drawn in this work, it will be neglected here for simplicity.

Once it is initialized, the hydrodynamic phase evolves in the usual way and is terminated when the system has cooled to a freeze-out temperature of $T_{\mathrm{fo}} = 120$ MeV, which is a typical value (cf., e.g., \cite{Ahmad:2016ods}).  This is done in lieu of terminating at a higher temperature and evolving subsequently with a hadron cascade.  After freeze-out, particle yields are obtained by evaluating Eq.~\eqref{CooperFrye} numerically as an integral over the freeze-out surface.

The correlation functions \eqref{C_q_K_vs_S_x_K_definition_woSA}-\eqref{C_q_K_vs_S_x_K_definition_wSA} can be similarly evaluated in terms of Cooper-Frye-like integrals over the freeze-out surface \cite{Plumberg:2016sig}.  In this case, the correlation function \eqref{C_q_K_vs_S_x_K_definition_woSA} is first evaluated on a fixed grid of points in $\vec{q}$, $K_T$, and the transverse pair momentum angle $\Phi_K$.  For each $K_T$ and $\Phi_K$, it is then fit to Eq.~\eqref{fitCF}, which gives a set of $R^2_{ij}$ as functions of $K_T$ and $\Phi_K$.  More details of the fitting procedure are described in \cite{Plumberg:2016sig}.  No systematic (e.g., fit-range \cite{Frodermann:2006sp}) uncertainties have been assessed for the fits in this study.

Once the fit radii are obtained, they are averaged separately over their angular dependence, finally yielding them as functions of $K_T$ only.  Since the radii are azimuthally averaged, they are basically insensitive to differences between the $x$ and $y$ directions in large systems which tend to change with centrality \cite{Plumberg:2015mxa}.

This highly simplified hydrodynamic model allows us to concentrate on the essential features of interest in this study, namely, the connection between the presence of enhanced flow in small systems and the resultant scaling of the $R^2_{ij}$ with multiplicity.

\section{Experimental context}
\label{Sec3}

\begin{figure*}
\centering
	\includegraphics[width=\textwidth, keepaspectratio]{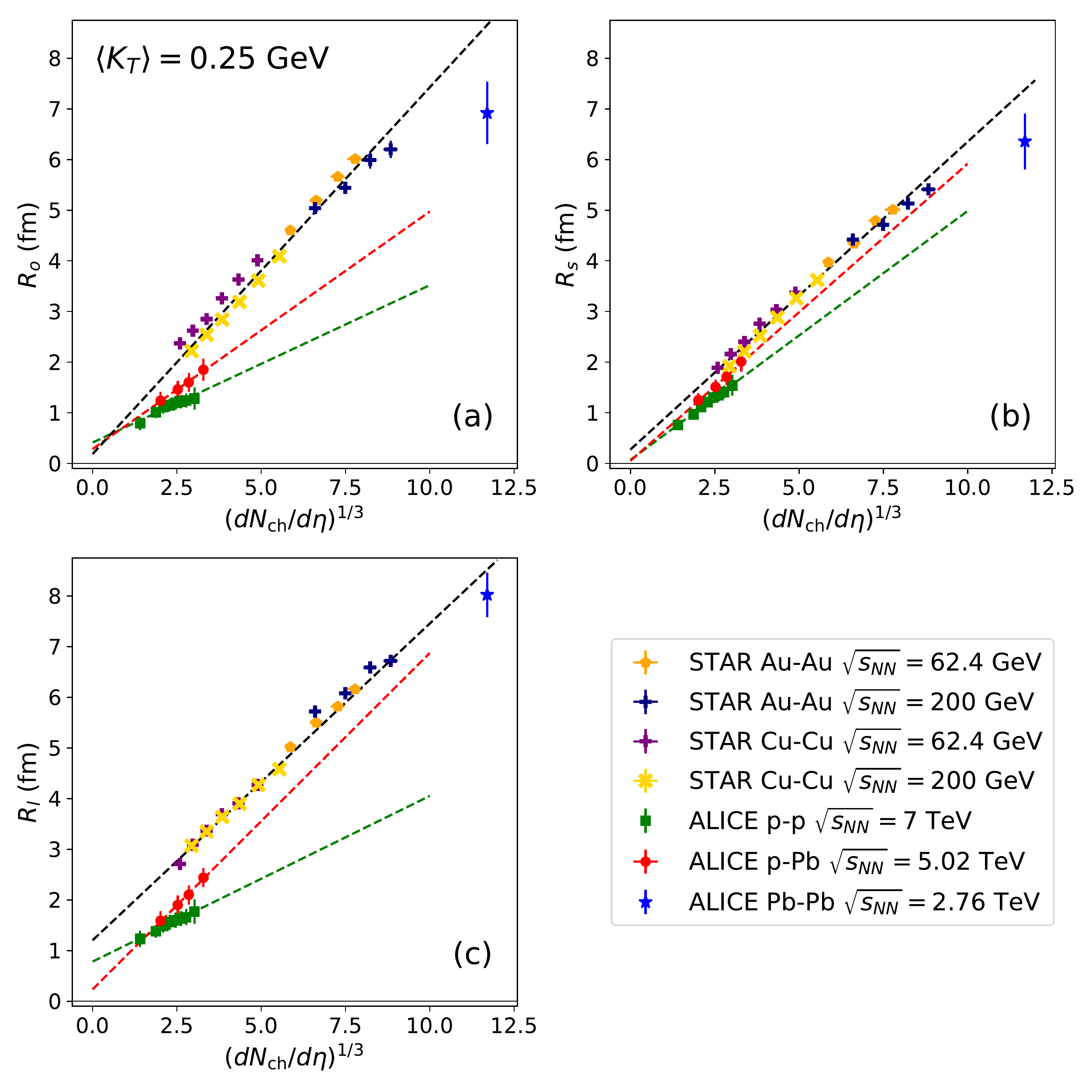}
	\caption{World data for several different femtoscopic analyses of both large and small systems.  In the slopes of the HBT radii vs. $(dN_\ch/d\eta)^{1/3}$, one observes both \textit{non-universality} (radii have different slopes in large vs. small systems) and \textit{hierarchy} (different radii possess differing slopes in a given system).  The STAR results were published in \cite{Adams:2004yc, Abelev:2009tp}.  The ALICE results for \pp, \pPb, and \PbPb were given in Refs.~\cite{Aamodt:2011kd, Adam:2015pya, Aamodt:2011mr}.  The fit lines were added by the author to guide the eye.  See the text for further discussion. \label{Fig:WorldData}}
\end{figure*}

The experimental motivation for this study originated from two analyses alluded to previously which explored the multiplicity dependence of HBT in \pp and \pPb collisions \cite{Aamodt:2011kd, Adam:2015pya} and compared it with similar measurements for larger systems (such as Au+Au \cite{Adams:2004yc, Abelev:2009tp} and \PbPb \cite{Aamodt:2011mr}).  Several of these measurements are shown in Fig.\ref{Fig:WorldData}.

There are two noteworthy features in this data which have been already discussed extensively in the literature \cite{Kisiel:2011jg, Aamodt:2011kd, Adamczyk:2014mxp, Adam:2015pya, Sirunyan:2017ies, Sikler:2017mde} and which were briefly described in the Introduction.  First, the radii across different systems exhibit discontinuities in their \textit{magnitudes}: that is, the values of the radii appear to differ across collision systems at fixed multiplicity.  Second, and somewhat related to the first point, the radii exhibit different \textit{slopes} with $(dN_\ch/d\eta)^{1/3}$ in the various directions and collision systems.  Fit lines (dashed) have been included in Fig.~\ref{Fig:WorldData} for \pp, \pA, and \AA, in order to guide the eye.\footnote{The \PbPb points in Fig.~\ref{Fig:WorldData} are all noticeably below the dashed trend lines of the STAR datasets.  This may be an effect of $K_T$ scaling \cite{Makhlin:1987gm}: the \PbPb point is in the 0-5\% centrality class with $K_T=0.2-0.3$ GeV, whereas the STAR points have $K_T=0.15-0.25$ GeV, meaning that the \PbPb point should fall somewhat below the STAR trends.  It is also possible that the discrepancy is affected by how the different centrality classes were determined: in the \pPb \cite{Adam:2015pya} and \PbPb \cite{Aamodt:2011mr} analyses, centrality was based on the signal in V0 forward/backward detectors, whereas in the STAR and \pp datasets, centrality was determined from multiplicities at mid-rapidity \cite{Adams:2004yc, Abelev:2009tp, Aamodt:2011kd}.

I will not try to sort out these issues out here, but will assume for simplicity that the \PbPb and STAR points all obey the same universal scaling behavior.  This assumption of a single \AA scaling can of course be revisited when the rest of the \PbPb centrality dependence is made available \cite{Aamodt:2011mr}.}  The axes in each panel have also been fixed to the same ranges, in order to facilitate the comparison of slopes in different radii.

The features noted in the Introduction -- the hierarchy and non-universality of the slopes -- can then be easily recognized in Fig.~\ref{Fig:WorldData}.  Non-universality is reflected in a comparison of the slopes of \textit{different} dashed lines in the \textit{same} panel; thus, in $R_o$ (panel (a)), \pPb has a smaller slope than \AA, while \pp is smaller than both.  Similarly, the slope hierarchy is manifest in comparing the \textit{same} datasets in \textit{different} panels; for instance, the \pp datasets (green squares with green dashed line) have different slopes in $R_o$ (1a), $R_s$ (1b), and $R_l$ (1c).  A more thorough inspection of the complete datasets in Fig.~\ref{Fig:WorldData} shows that these features also depend strongly on the $K_T$ value for which they are plotted (this will be seen clearly below in Sec.~\ref{Sec4}).

It is important to emphasize here that the hierarchy and non-universality of the slopes in Fig.~\ref{Fig:WorldData} are entirely independent concepts: one could have had completely universal slopes which exhibited a hierarchy (i.e., were independent of collision system, but differed for each radius), or one could as easily have had no hierarchy between the various radii, but a non-universal slope for each radius whose value depended on the collision system.  In the present case, of course, a mixture of both is found in the data of Fig.~\ref{Fig:WorldData}.  One finds in \AA, for instance, that the slope of $R_o$ is somewhat larger ($\sim 0.7$) than that of $R_s$ or $R_l$ ($\sim 0.6$), whereas for \pp, the trend is reversed: the $R_o$ fit has a slope comparable to $R_l$ ($\sim 0.3$), while the $R_s$ slope is considerably steeper ($\sim 0.5$).  This paper is an attempt to organize these various observations within a single, coherent framework.


As already noted, many other works have already observed the discrepant behavior in $R_o$ when compared with $R_s$ and $R_l$ \cite{Kisiel:2011jg}, although these observations have sometimes been made only for larger collision systems \cite{Lisa:2005dd}.  Previous theoretical work has explored the implications of the $K_T$-dependence of the radii \cite{Hirono:2014dda} but has not specifically considered the role of the $(dN_\ch/d\eta)^{1/3}$ dependence in the radii.  Refs.~\cite{Graef:2012sh,Graef:2012za,Li:2012np} have further emphasized the importance of flow and the space-time structure of the source for understanding the $(dN_\ch/d\eta)^{1/3}$ dependence of the \pp and \PbPb radii as modeled by UrQMD, but do not appear to have analyzed the same dependence in detail from the perspective of hydrodynamics.

For the present study, the goal is to explore specifically whether the differences in the $(dN_\ch/d\eta)^{1/3}$ dependence between the various radii and collision systems can be naturally understood in terms of the space-time picture provided by hydrodynamics.  Since the hydrodynamic formalism used here is highly simplified in the interest of clarity, the focus will be placed on obtaining a qualitatively plausible understanding of how hydrodynamics describes the space-time evolution in different collision systems, rather than attempting to quantitatively reproduce the data in detail.  This is the subject to which we turn next.

\section{Results}
\label{Sec4}

In this work, I have applied the formalism covered in Sec.~\ref{Sec2} to the systems studied in the experimental analyses described in Sec.~\ref{Sec3}.  The results are presented in this section and are organized into three areas.  In order to understand the multiplicity scaling of the HBT radii, we must first appreciate the ways in which changing the multiplicity in different systems affects the shape of the freeze-out surfaces themselves whose structure the HBT radii are supposed to characterize.  We begin by examining and comparing these surfaces directly in Fig.~\ref{Fig:FOsurfaces} for different systems and centrality classes.  This will lead us to consider how this space-time structure can be manifested in the HBT radii, and it is at this point that the Gaussian source approximation will prove useful for guiding intuition, namely, by relating the HBT radii directly to space-time variances of the source function.  Finally, having an intuitive feeling for how the multiplicity influences HBT on the basis of hydrodynamics, we finally consider the radii themselves which are extracted according to the above formalism.

\subsection{Freeze-out surfaces}

\begin{figure*}
\centering
\includegraphics[width=\textwidth, keepaspectratio]{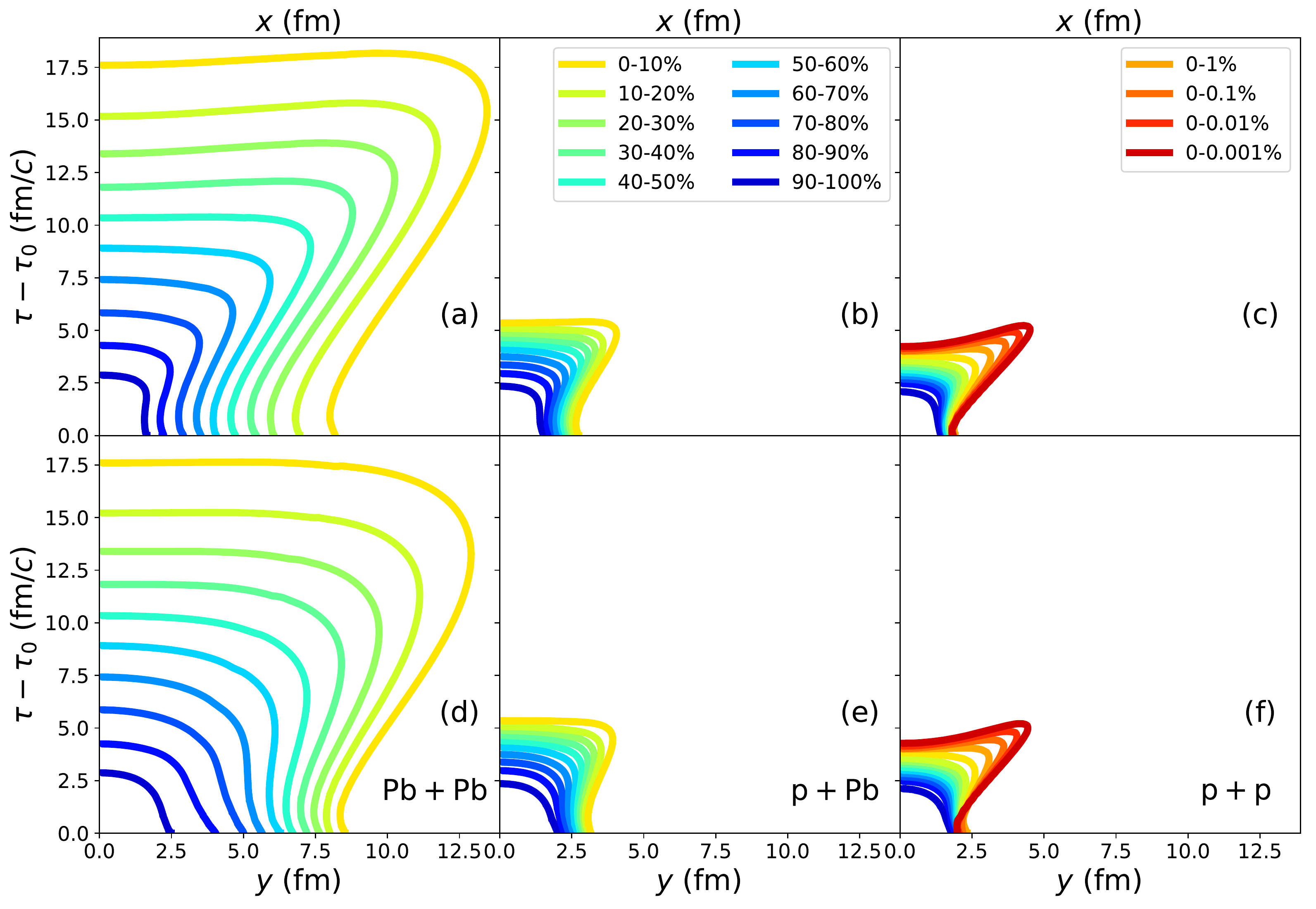}
\caption{Freeze-out surfaces for various centrality classes in (from right to left) \pp, \pPb, and \PbPb collisions.  While increasing the multiplicity proportionately increases the size of the system, it also tends to distort the shape of the freeze-out surface.  Several additional centrality classes are shown for \pp, in order to illustrate the changes in shape which occur at high multiplicity.  \label{Fig:FOsurfaces}}
\end{figure*}

First, since HBT interferometry reflects the space-time structure of the emitting source in nuclear collisions, it is crucial to examine how the freeze-out surfaces themselves evolve with the system's multiplicity.  This is shown in Fig.~\ref{Fig:FOsurfaces} for the centrality classes under consideration; similar plots were also studied in Ref.~\cite{Heinz:2019dbd}.  Several additional classes for \pp have also been added at large multiplicities for illustrative clarity.  One notices immediately a conspicuous difference in the behavior of \PbPb collisions as compared with \pPb and \pp collisions, especially at large centralities.  In the former, the scaling with multiplicity primarily affects only the enclosed space-time volume of the system, without dramatically altering the shape of the freeze-out surface itself.  In small systems, however, the growth of the enclosed volume with multiplicity is less important than changes to the shape of the freeze-out surface, especially at high multiplicity.  Remarkably, in extreme \pp collisions, the system freezes out in the center \textit{first}, followed by freeze-out at the edges.  Viewed as an animation, one would see such a system as a ring of quark-gluon plasma in the transverse plane, expanding and narrowing until final freeze out at a time $\tau - \tau_0 \sim 5$ fm$/c$ and radius $r \sim 4.5$ fm.  One should therefore expect significant differences in the scaling of the radii in \PbPb collisions when compared to that in \pp or \pPb collisions at similar multiplicities.

One also notices, by comparing the slices along the $x$ and $y$ axes, that \pp and \pPb collisions exhibit greater rotational symmetry than \PbPb: this reflects the fact that the increasing importance of event-by-event fluctuations in the location and violence of collisions between subconstituents destroys the rather tight correlation of collision centrality and impact parameter observed in collisions between large nuclei when going to small collision systems \cite{Adam:2014qja, Welsh:2016siu}.  It is also worth underscoring that the contours in Fig.~\ref{Fig:FOsurfaces} correspond to fixed centrality classes, not necessarily fixed multiplicities.  The comparison at fixed multiplicity will be shown below.


From these reflections we may already draw a very important preliminary conclusion: hydrodynamics does not in general predict a universal scaling of the $R^2_{ij}$ with $dN_\ch/d\eta$ which is irrespective of the system size.  Conversely, even highly simplified hydrodynamic models (like the one considered here) predict a non-universal scaling for high-multiplicity \pp collisions.  Although this observation is focused on HBT and a particular definition of the multiplicity, it presumably applies to other space-time observables and definitions of the multiplicity as well.

\subsection{The emission function $S(x,K)$}

\begin{figure*}
\centering
\includegraphics[width=\linewidth, keepaspectratio]{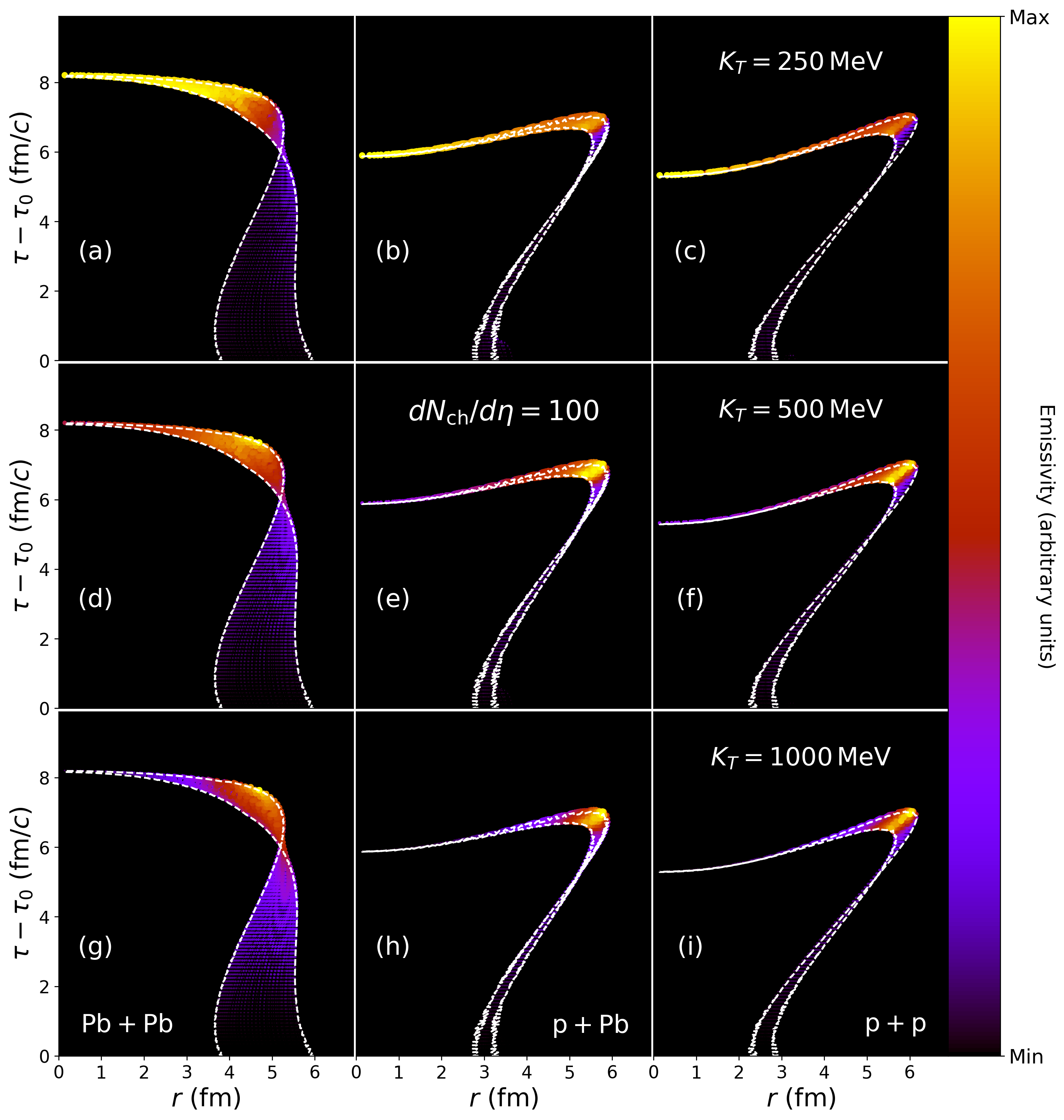}
\caption{The emission function plotted for various $K_T$ as a brightness density distribution over the freeze-out surfaces in \pp, \pPb, and \PbPb (right to left).  The brightest regions (yellow) correspond to fluid cells emitting the largest fraction of pions at the given value of $K_T$; similarly, the darkest (purple, black) points represent the cells which make little to no contribution to the final pion yield.  The initial conditions for \pp have been artificially rescaled so that the minimum bias $dN_\ch/d\eta = 100$, while the \pPb and \PbPb surfaces were obtained for collisions of 54.9-64.9\% and 0-0.00025\%, respectively, with the normalizations determined from Table \ref{Table1}.  \label{Fig:emission_densities}}
\end{figure*}

The fact that the freeze-out structure scales differently in large than small collisions should be reflected in the radii as well.  This can be justified quantitatively by considering the scaling of the source variances entering the $R^2_{ij}$ that are obtained using the Gaussian source approximation.  Recall that, using this approximation, the HBT radii can be related directly to the space-time structure of the underlying emission function.  The multiplicity dependence of the HBT radii is therefore approximately reflected in the corresponding behavior of the brightest emission regions at a fixed value of $K_T$.

To see this more clearly, we write out explicitly the $R^2_{ij}$ of interest on the basis of Eq.~\eqref{GSA_rel1}:

\begin{eqnarray}
R^2_s &=& \l< \tilde{x}_s^2 \r> \label{R2s_GSA}\\
R^2_o &=& \l< \tilde{x}_o^2 \r>
          - 2\beta_T\l< \tilde{x}_o \tilde{t} \r>
          + \beta_T^2\l< \tilde{t}^2 \r> \label{R2o_GSA}\\
R^2_l &=& \l< \tilde{x}_l^2 \r>
          - 2\beta_L\l< \tilde{x}_l \tilde{t} \r>
          + \beta_L^2\l< \tilde{t}^2 \r> \label{R2l_GSA}
\end{eqnarray}

These relations depend on a total of six source variances: three geometric terms ($\l<\tilde{x}_o^2\r>$, $\l<\tilde{x}_s^2\r>$, $\l<\tilde{x}_l^2\r>$), two cross terms ($\l<\tilde{x}_o \tilde{t}\r>$, $\l<\tilde{x}_l \tilde{t}\r>$), and a purely temporal term $\l<\tilde{t}^2\r>$.  Each term probes a different spatiotemporal dimension of the effective emission region which dominates particle production for a given $\vec{K}$.  Thus, for instance, $\sqrt{\l<\tilde{x}_o^2\r>}$ represents the size of a given emission region in the out direction (along the direction of $\vec{K}$ in the transverse plane).  Similarly, $\sqrt{\l<\tilde{t}^2\r>}$ represents the spread in times over which particles at a given $\vec{K}$ were typically emitted, while the $\l<\tilde{x}_o \tilde{t}\r>$ represents the degree of correlation between the out and time coordinates of particle emission.

Now we wish to see how the scaling of these source variances is reflected in the emission regions shown in Fig.~\ref{Fig:emission_densities}, which compares \pp, \pPb, and \PbPb collisions at a fixed $dN_\ch/d\eta=100$ as was done in Ref.~\cite{Heinz:2019dbd}.  Note that, as was observed in \cite{Heinz:2019dbd}, the condition of equal multiplicity requires each freeze-out surface portrayed in Fig.~\ref{Fig:FOsurfaces} to have the same co-moving volume; visual inspection shows that this condition is very closely satisfied.  In addition, however, the color map in Fig.~\ref{Fig:emission_densities} projects the emission function $S(x,K)$ for each system directly onto its respective freeze-out surface, in order to illustrate how the space-time structure of $S$ is influenced by $K_T$ and the collision system under consideration.\footnote{Note that, in the case of \pp, the normalization of the minimum-bias initial conditions was retuned in order to reach $dN_\ch/d\eta = 100$, since these events are too rare to be conveniently reproduced with the normalization fixed by Table \ref{Table1}.  This is only a justifiable trick in \pp where the multiplicity is not correlated with the impact parameter $b$ the way it is in larger systems, so that the rescaling does not alter the subsequent evolution significantly.  In any event, the retuning to $dN_\ch/d\eta=100$ is used here for purely illustrative purposes, in order to indicate how different systems may possess dramatically different space-time structures, even at the same multiplicity.}

In each panel of Fig.~\ref{Fig:emission_densities}, the outlines of the freeze-out surface along the $x$ and $y$ axes are shown as thin, dashed white lines to guide the eye.  Bright yellow regions have the highest pion emissivity and will tend to dominate the corresponding source variances as well; dark purple regions contain fluid cells producing the fewest pions and will accordingly have little effect on the source variances.  The color scale is normalized between 0 (minimum emissivity) and 1 (maximum emissivity).


Thus, one observes that at small $K_T$, particle emission happens mostly at late times $\tau \sim 5-8$ fm$/c$ within $2-4$ fm of the origin ($r=0$).  In \PbPb, emission at small $K_T$ occurs later and faster than emission at large $K_T$, which originates mainly from the edge of the system over a larger and earlier spread of times.  In \pp and \pPb, by contrast, small $K_T$ emission happens \textit{earlier} and more rapidly than large $K_T$ emission, as a consequence of the systems' freezing out sooner in the centers than at the edges.  Adjusting the $K_T$ window thus influences where the emission function is brightest, and thereby provides a tunable filter with which to probe different portions of the freeze-out surface in a controlled way.

The shifting of the highest emissivity regions with $K_T$ leads to the well-known $K_T$ scaling of the radii \cite{Makhlin:1987gm, Lisa:2005dd} which can be identified directly in the reduced sizes of the bright yellow regions at large vs. small $K_T$ in Fig.~\ref{Fig:emission_densities}.  This effect is present in all three systems and is a consequence of collective flow: particles emitted from the system's center tend to belong to pairs with relatively small $K_T$ values, since the collective motion is comparatively weak there.  Particles emitted from the system's edge, on the other hand, are produced by fluid elements which already possess a strong transverse velocity component, and consequently emit particles preferentially with large momenta moving in the same direction.  Large-$K_T$ emission is thus dominated by the fluid cells at the edge of the system, leading to more compact emission regions, and causing the HBT radii to decrease accordingly \cite{Heinz:2004qz}.

Moreover, the highly elongated `wing-like' structure of the small systems' freeze-out surfaces is also a result of enhanced collective flow, in which the center of the source freezes out well before the edges do.  In this sense, small systems at high multiplicity are quite literally exploding `rings of fire,' from the perspective of hydrodynamics.  This enhanced collective flow in small systems originates from a combination of their reduced sizes (generating larger initial density gradients) and the higher temperatures produced in their interiors \cite{Sievert:2019zjr} which generate a more violent response due to a larger speed of sound (e.g., \cite{Borsanyi:2013bia, Bazavov:2014pvz}).

\begin{figure*}
\centering
\includegraphics[width=0.5\linewidth, keepaspectratio]{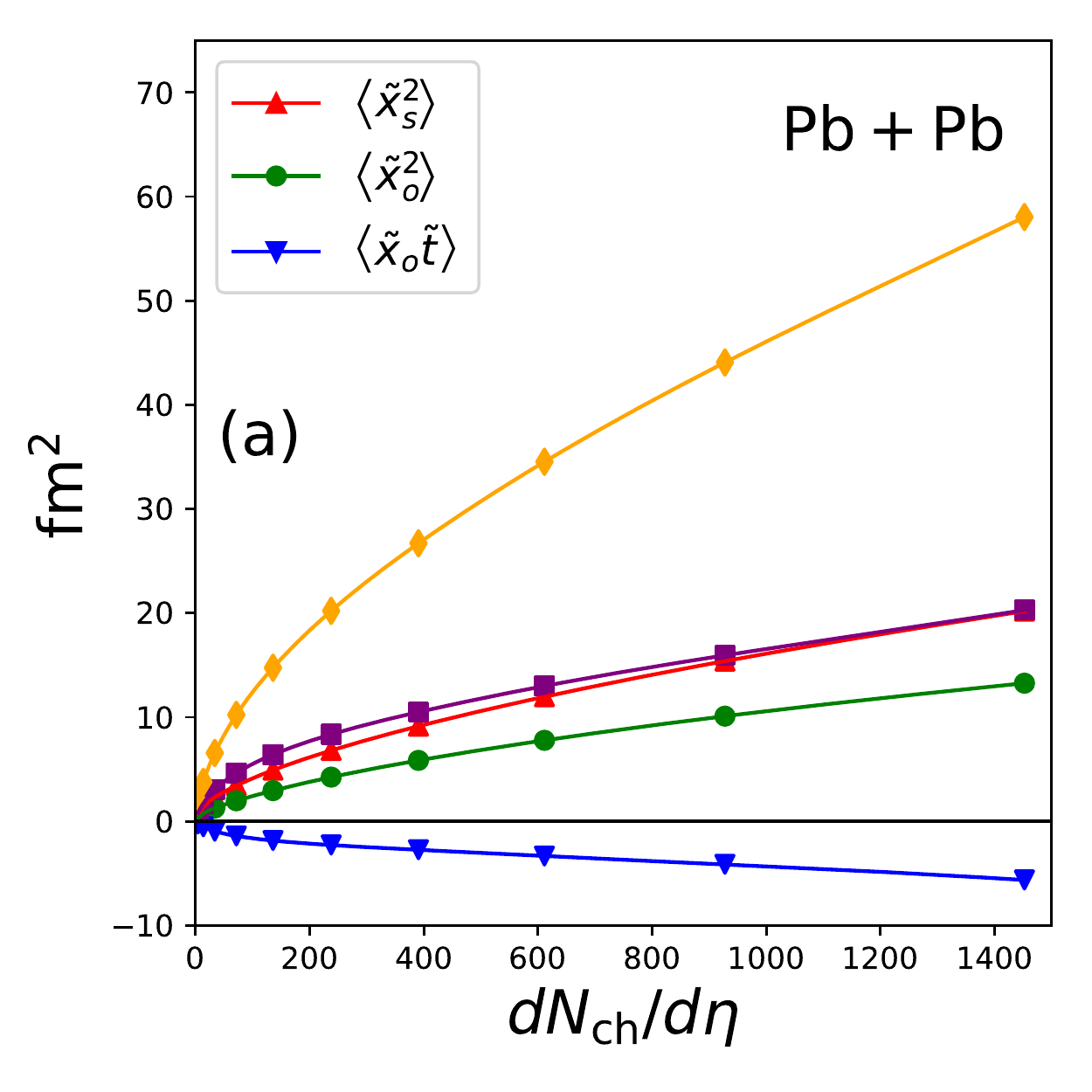}%
\includegraphics[width=0.5\linewidth, keepaspectratio]{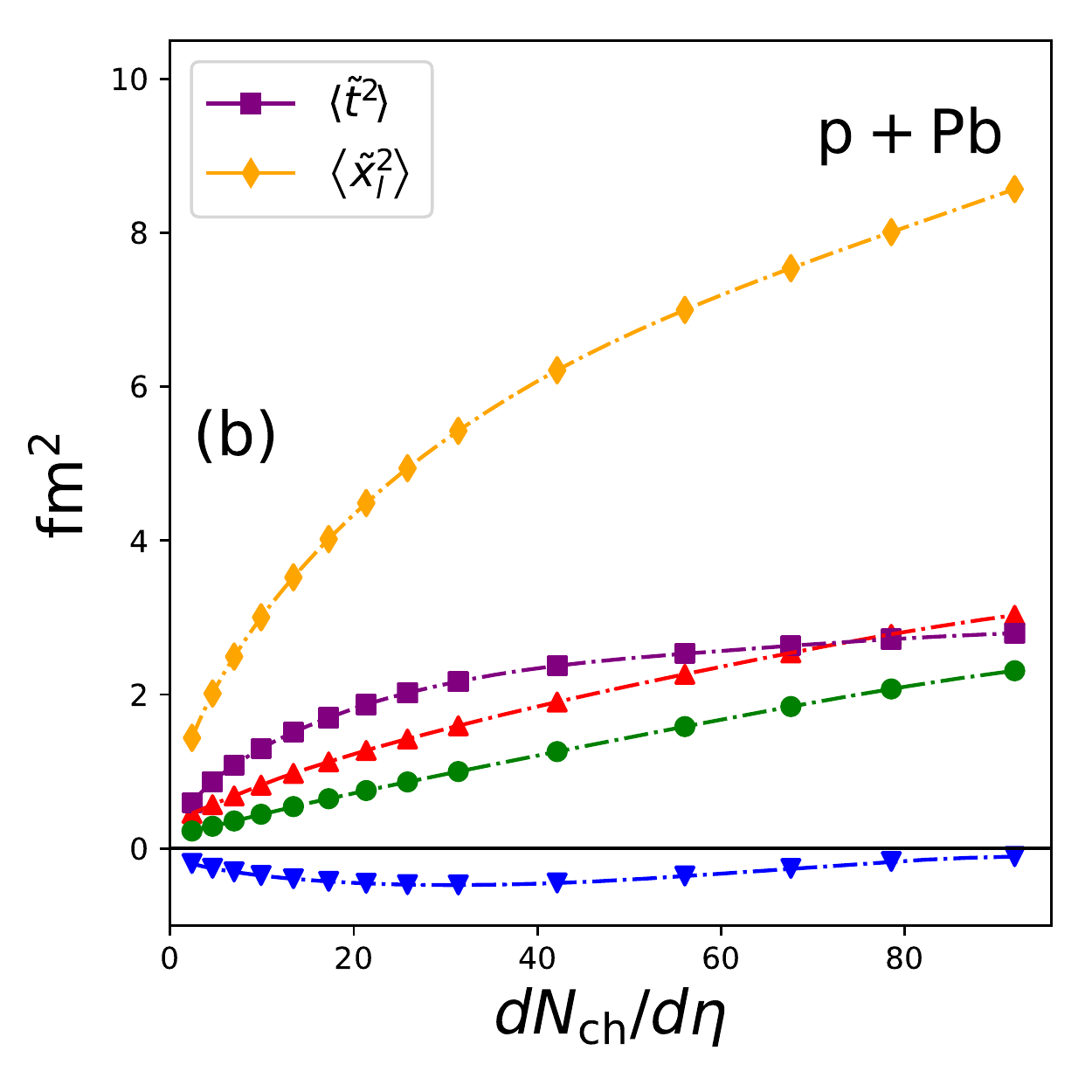}\\
\includegraphics[width=0.5\linewidth, keepaspectratio]{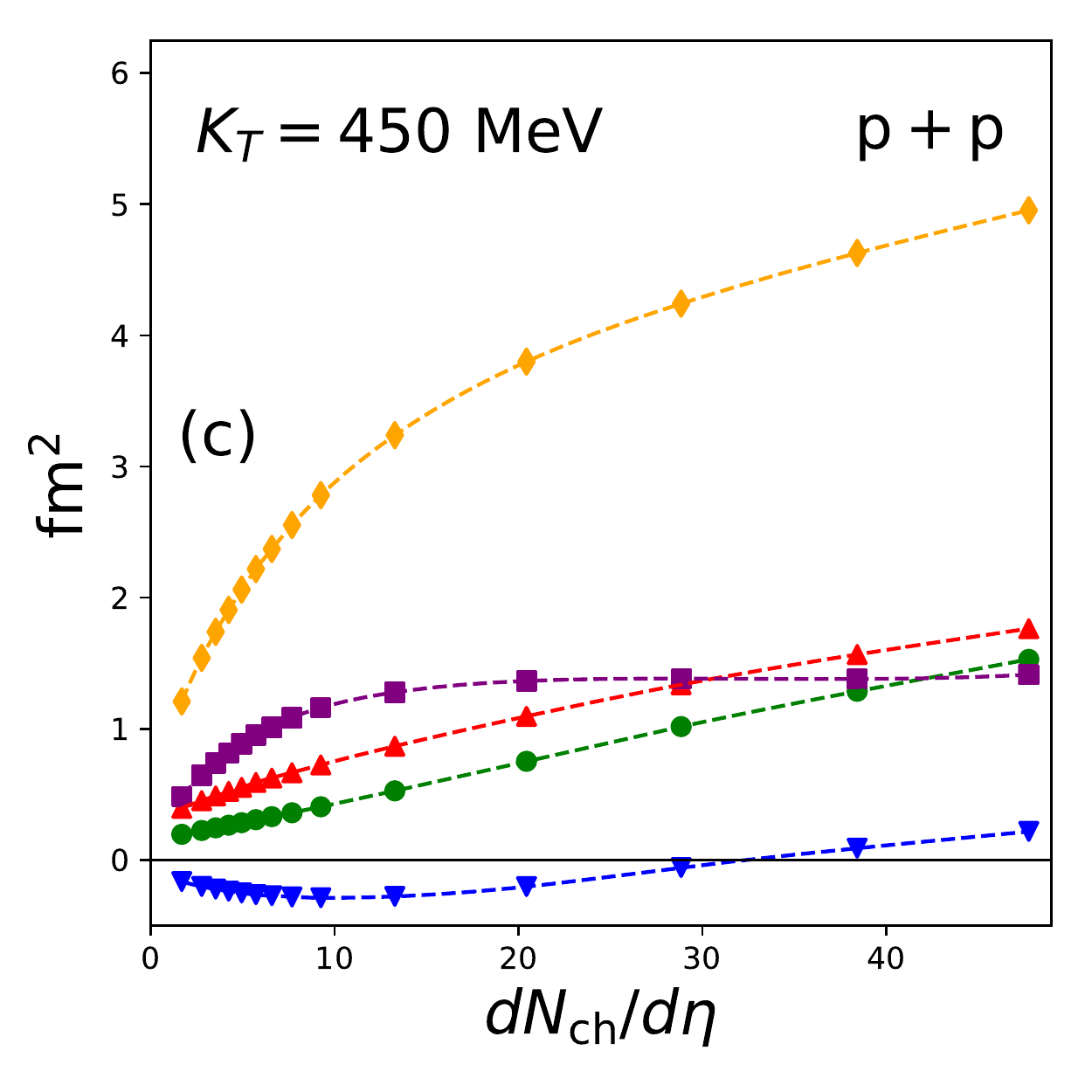}%
\includegraphics[width=0.5\linewidth, keepaspectratio]{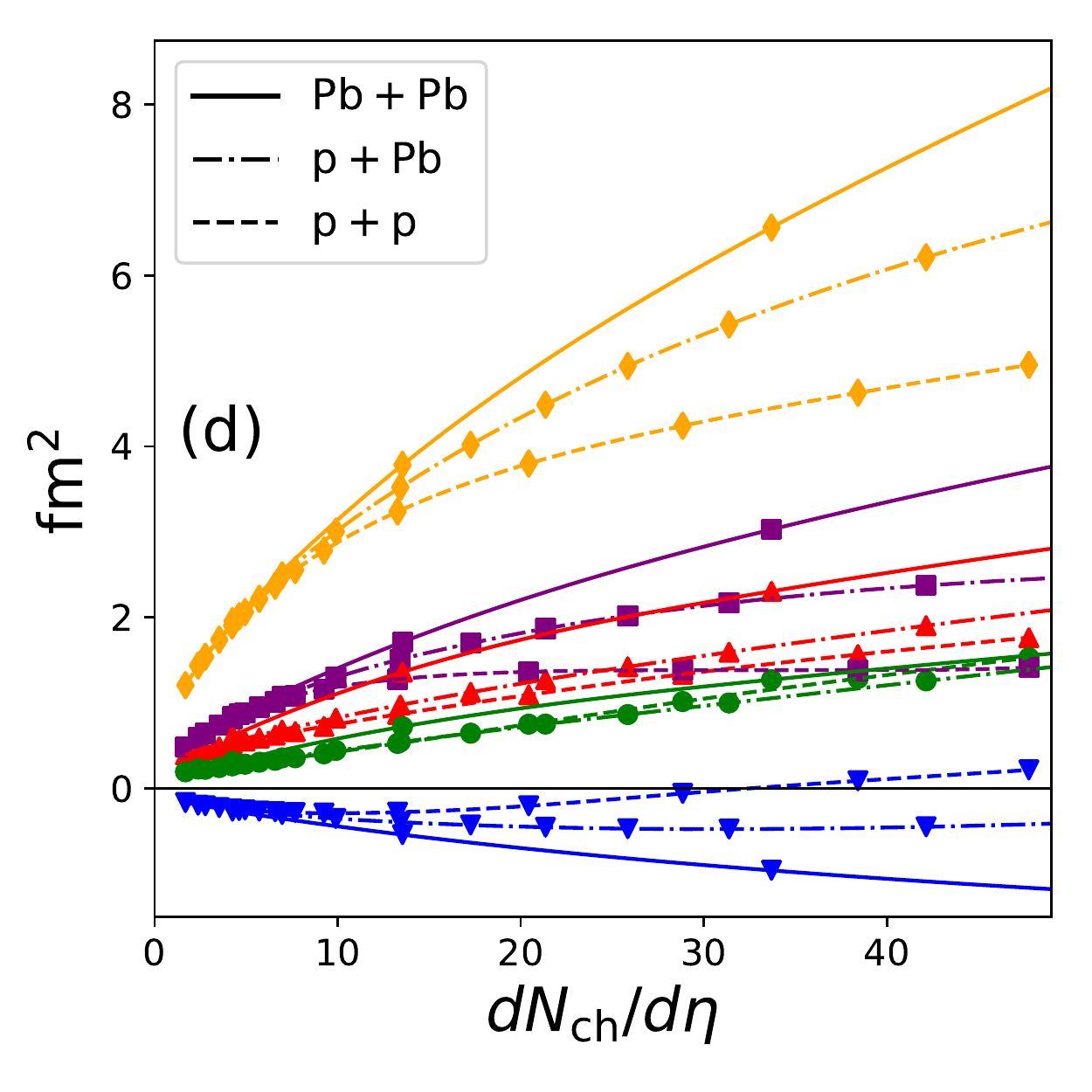}
\caption{The source variances entering $R^2_o$, $R^2_s$, and $R^2_l$ in the Gaussian source approximation (cf. Eqs.~\eqref{R2s_GSA}-\eqref{R2l_GSA}) as functions of $dN_\ch/d\eta$ in \PbPb [panel (a)], \pPb [panel (b)], and \PbPb [panel (c)].  The panels (a-c) are compared side-by-side in panel (d) on the same scale, in order to make the differences between systems more apparent.  Here, $K_T = 450$ MeV.  One observes that most of the splitting between systems emerges in the temporal or longitudinal variances ($\l< \tilde{x}_l^2 \r>$, $\l< \tilde{x}_o \tilde{t} \r>$, $\l< \tilde{t}^2 \r>$), whereas less splitting is visible in the transverse geometry ($\l< \tilde{x}_o^2 \r>$, $\l< \tilde{x}_s^2 \r>$).  All source variances have been averaged azimuthally over $\Phi_K$. \label{Fig:SVs_vs_dNchdeta}}
\end{figure*}

Notably, this elongated freeze-out structure leads visibly at large $K_T$ to a strong, \textit{positive} correlation between $\tilde{x}_o$ ($\sim r$) and $\tilde{t}$, implying that $\l< \tilde{x}_o \tilde{t} \r> > 0$ in small systems at high multiplicity.  This is opposite to the behavior of \textit{large} systems at the same multiplicity in hydrodynamics, which clearly tend to have $\tilde{x}_o$ and $\tilde{t}$ \textit{negatively} correlated with one another, implying that $\l< \tilde{x}_o \tilde{t} \r> < 0$ in these systems.

The freeze-out geometry also has implications for the scaling of the other source variances with multiplicity, as shown in Fig.~\ref{Fig:SVs_vs_dNchdeta} for \pp, \pPb, and \PbPb at $K_T=450$ MeV.  Since we work here in the LCMS frame \cite{Csorgo:1991ej, Chapman:1994ax}, in which the longitudinal component of the pair momentum vanishes ($K_L \equiv 0$) for each pair by definition, $\beta_L = 0$ as well, so that both $R^2_s$ and $R^2_l$ are dominated completely by the system's spatial geometry.  The geometric variances ($\l< \tilde{x}_o^2\r>$, $\l< \tilde{x}_s^2\r>$, $\l< \tilde{x}_l^2\r>$) shown in Fig.~\ref{Fig:SVs_vs_dNchdeta} grow approximately monotonically with multiplicity in both large and small systems, reflecting the steady scaling which is already visible in Fig.~\ref{Fig:FOsurfaces}.  $\l< \tilde{x}_s^2\r>$ (red up-triangles) is found to be consistently larger than $\l< \tilde{x}_o^2\r>$ (green circles) in all systems, but both scale with $dN_\ch/d\eta$ in essentially the same way.  $\l< \tilde{x}_l^2\r>$ (orange diamonds) grows more rapidly than $\l< \tilde{x}_o^2\r>$ or $\l< \tilde{x}_s^2\r>$, reflecting the extended shape of the system in the longitudinal direction.

The behavior of the temporal variances ($\l< \tilde{x}_o \tilde{t} \r>$, $\l<  \tilde{t}^2\r>$) is somewhat more interesting.  In \PbPb collisions one observes that the emission duration $\l<  \tilde{t}^2\r>$ (purple squares) grows monotonically with multiplicity, while the correlation term $\l< \tilde{x}_o \tilde{t} \r> < 0$ everywhere and decreases monotonically with multiplicity (blue down-triangles).  In \pp and \pPb, on the other hand, this monotonic behavior is lost: the emission duration eventually ``levels off" and the correlation term actually turns positive $\l< \tilde{x}_o \tilde{t} \r> > 0$ at a critical value of the multiplicity, owing to the wing-like structure shown already in Fig.~\ref{Fig:emission_densities}.

In Fig.~\ref{Fig:SVs_vs_dNchdeta}d, we show the same source variances as in panels (a-c), but overlaid on the same set of axes to facilitate direct comparison.  We observe a number of critical features.  First, the longitudinal variance scales strongly with the size of the collision system, with noticeable splitting occurring above $dN_\ch/d\eta \gtrsim 12$.  Similar splitting is seen in $\l< \tilde{x}_o \tilde{t} \r>$ and $\l< \tilde{t}^2 \r>$, for which the freeze-out geometry dictates dramatically different behavior in the different systems, as we have seen previously.  Interestingly, the transverse source variances $\l<x_s^2\r>$ and, in particular, $\l<x_o^2\r>$ change surprisingly little between large and small systems when holding the multiplicity fixed. At a superficial level this reflects the observation made in \cite{Heinz:2019dbd} that, if freeze-out occurs at constant density, for fixed multiplicity the co-moving volume must be the same in all collision systems. However, HBT radii are known not to measure the entire freeze-out volume, but only some fraction of it, known as the `homogeneity volume' \cite{Akkelin:1995gh}, whose size is affected by the collective expansion rate at freeze-out. As already discussed and explicitly seen in Figs.~\ref{Fig:FOsurfaces} and \ref{Fig:emission_densities}, this expansion rate increases from \PbPb to \pp collisions; Fig.~\ref{Fig:SVs_vs_dNchdeta}d shows that, at fixed multiplicity this increase in radial flow leads to a decrease of $\l<x_s^2\r>$ from \PbPb (solid) to \pPb (dash-dotted) to \pp (short-dashed), as anticipated in \cite{Heinz:2019dbd}. Contrary to the expectations in \cite{Heinz:2019dbd}, however, this effect is much weaker for $\l<x_o^2\r>$ and can therefore not explain the experimentally observed significantly larger variation of $R_o^2$ than $R_s^2$ when going from \PbPb to \pp at fixed multiplicity.  Instead, as we will discuss next, this last feature can be understood by studying the system size dependence of the other contributions to $R_o^2$ in Eq.~\eqref{R2o_GSA}, caused by the qualitative change in the shape of the freeze-out surface exhibited in Fig.~\ref{Fig:emission_densities}.

\subsection{The HBT radii $R^2_{ij}$}

\begin{figure*}[!t]
\centering
\includegraphics[width=\textwidth, keepaspectratio]{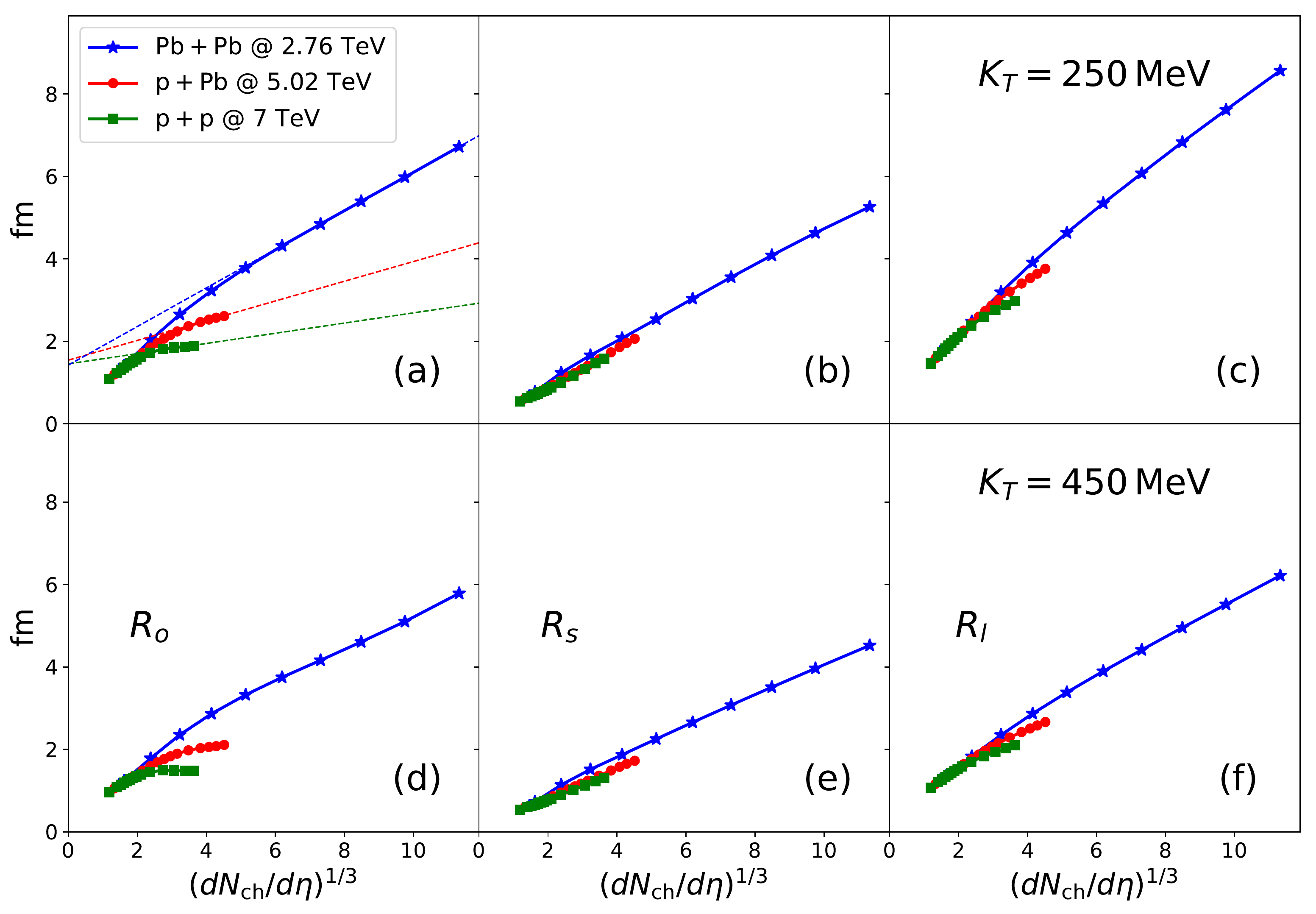}\\
\caption{The multiplicity scaling of the 3D HBT radii in a simplified hydrodynamic model of \pp at 7 TeV, \pPb at 5.02 TeV, and \PbPb at 2.76 TeV.  Results for $K_T = 250$ MeV are compared with those for $K_T = 450$ MeV.  In panel (b), some illustrative fits to the high-multiplicity trends of $R_o$ are included to accentuate the splitting in the slopes.  See the text for discussion.\label{Fig:R2ij_vs_dNchdeta}}
\end{figure*}

We are finally in a position to consider the actual multiplicity dependence of the HBT radii in hydrodynamics.  This is shown in Fig.~\ref{Fig:R2ij_vs_dNchdeta}.  As expected, $R_s$ and $R_l$ follow a nearly universal, approximately linear scaling with $(dN_\ch/d\eta)^{1/3}$ in all three systems.  Since they are dominated by the spatial geometry of the system, their scaling reflects the extensive nature of $dN_\ch/d\eta$, which should be proportional to the system's volume.

$R_o$, on the other hand, is sensitive to both the spatial and temporal sizes of the source, as well as the correlation between the two.  As we have just seen, the latter behaves very differently in large and small collision systems at a fixed multiplicity: the wing-like geometry induced by strong collective flow forces a change of sign in the correlation term $\l< \tilde{x}_o \tilde{t} \r>$ and a concomitant leveling off of the emission duration $\l< \tilde{t}^2 \r>$.  The combination of these effects is that, in high-multiplicity \pp and \pPb collisions, $R_o$ exhibits a much weaker scaling with multiplicity than either $R_s$ or $R_l$, whereas in \PbPb collisions, the comparatively weaker flow allows $R_o$ to grow at a rate similar to that seen in $R_s$ and $R_l$.  The shallow scaling of $R_o$ with $(dN_\ch/d\eta)^{1/3}$ in \pp and \pPb becomes especially pronounced at higher multiplicities, leading to a slower overall growth with multiplicity.  This is how hydrodynamics explains the slope hierarchy observed in the data.

Hydrodynamics may also allow an understanding of the slope non-universality visible across collision systems.  Hydrodynamics predicts very similar slopes for $R_s$ in all systems, a feature which seems to be fairly well borne out by the data (cf.~panel (b) of Fig.~\ref{Fig:WorldData}).  For $R_o$, the slope in \pPb falls squarely in between those of \pp and \PbPb at large multiplicities, again in surprisingly good agreement with data.

The $R_l$ data initially seem to violate the qualitative hydrodynamic tendencies, showing similar slopes between \pPb and \PbPb, but a significantly smaller slope in \pp (cf.~panel (c) of Fig.~\ref{Fig:WorldData}).  On closer inspection, however, the discrepancies may not be as bad as they first appear: at large multiplicities, there is a small but detectable splitting in the model slopes of $R_l$ which mirrors that seen in $R_o$ and qualitatively, if not quantitatively, resembles the splitting seen in the data.  Although a thorough resolution of this tension is beyond the scope of the current work, it is worth speculating as to how the tension originates and how it might be alleviated.  The qualitative similarity may originate from the fact that both $\l< \tilde{z}^2 \r>$ and $\l< \tilde{t}^2 \r>$ are affected by the boost-invariant structure of the hydrodynamic model used here: the `leveling off' in the spread of emission times (cf.~panels (a) and (b) of Fig.~\ref{Fig:SVs_vs_dNchdeta}) would in this case lead to a similar, but much weaker leveling off in the spread of emission \textit{positions} in the longitudinal direction, which is manifested in a splitting of $R_l$.  The fact that the splitting is so much smaller in hydrodynamics than in the data is likely due to the fact that the LCMS condition ($K_L = 0$) is enforceable exactly in the model considered here, but when imposed experimentally requires adopting a slightly different Lorentz frame for each pair used in constructing the correlation function.  This effectively averages \eqref{2pCF_def} over the longitudinal and temporal properties of the underlying sources (cf.~\eqref{R2l_GSA} with $\beta_L \neq 0$) \cite{Chapman:1994ax}, and might explain the larger longitudinal slope hierarchy seen in the data.  Hydrodynamics will therefore tend to underestimate the splitting of $R_l$ until it is supplemented with more realistic features, such as a hadronic rescattering phase, which naturally produces a mixing of longitudinal and temporal source properties as is inherent to experimental analyses with finite statistics \cite{Chapman:1994ax}.  Of course, this proposed explanation is highly speculative and should be considered with appropriate caution.  For present purposes, it is sufficient to note that the model-to-data discrepancies seen between Figs.~\ref{Fig:WorldData}(c) and \ref{Fig:R2ij_vs_dNchdeta}(c, f) may be plausibly attributed to the simplicity of the model used here, and need not imply any intrinsic limitations of hydrodynamics itself.

There is another respect in which the model used here fails to completely represent the data.  It is clear that the multiplicity dependence of $R_o$ predicted by hydrodynamics is not linear in small systems and that there is even some slight curvature visible in the \PbPb curves shown here.  The reason for this is the competition between spatial and temporal information which influences $R_o$: because the freeze-out structure \textit{does} change with multiplicity (due to changes in the amount of flow) in both large and small systems, one expects these changes in shape to produce deviations from the otherwise linear dependence which would result from a pure rescaling of the system size.  Because all hydrodynamic systems change both size and shape with multiplicity (cf.~Fig.~\ref{Fig:FOsurfaces}), the scaling of the radii is not in general expected to be perfectly linear.

However, the fact remains that the current hydrodynamic model predicts non-linear $(dN_\ch/d\eta)^{1/3}$ dependence of the radii which is not obviously reflected in the data, and only reproduces the observed slope hierarchy and non-universality at multiplicities large enough to generate the wing-like structure of Fig.~\ref{Fig:emission_densities}.  Thus, one might worry that the $R_o$ scaling in small systems is too similar to the steeper $R_s$ and $R_l$ scaling at low multiplicities for the connection with hydrodynamics to be justifiably drawn in this regime.  While it remains to be seen whether a more sophisticated model would reproduce the trends observed in the data, it is worth pointing out that nothing in principle prevents hydrodynamics from being applicable even to systems with very low multiplicity \cite{Heinz:2019dbd}, and the more crucial question is whether sufficient flow can be generated to weaken the $R_o$ scaling also at smaller $dN_\ch/d\eta$ once effects like preequilibrium flow \cite{Hanus:2019fnc} have been included in the analysis.  This question will have to be answered with a more advanced model than the one employed here.

In the interest of clarity, it is helpful to present the data alongside the model results in a way which isolates the behavior of the slopes in the radii as themselves functions of $K_T$.  This can be done by estimating the slope (using simple linear fits) for a given radius in each system separately.  Fig.~\ref{Fig:R2ij_vs_dNchdeta} shows the $(dN_\ch/d\eta)^{1/3}$ dependence of the model results explicitly for only two values of $K_T$, and as already noted, in both cases it is clear that the differences in slopes only begin to emerge above a certain value of the multiplicity which depends on the system in question.  For \pp and \pPb shown in Fig.~\ref{Fig:R2ij_vs_dNchdeta}, this seems to happen for (roughly) the five largest centrality bins considered, corresponding to $dN_\ch/d\eta \gtrsim 13$ in \pp, and $dN_\ch/d\eta \gtrsim 42$ in \pPb.  \PbPb has a nearly constant slope for all multiplicities shown, although $R_o$ shows a slight curvature.  To isolate the slopes in these high-multiplicity regimes, we fit each radius against $(dN_\ch/d\eta)^{1/3}$ to a straight line over the five largest centralities and extract the corresponding slope; cf.~the illustrative fits included in Fig.~\ref{Fig:R2ij_vs_dNchdeta}a.  We perform a similar exercise for the experimental datasets presented in Fig.~\ref{Fig:WorldData}, using all published centrality or multiplicity datasets in the \pp, \pPb, and the combined STAR datasets.  For each radius and collision system in the model or data, we extract the approximate linear slope as just described and plot it as a function of $K_T$.

\begin{figure*}
\includegraphics[width=\linewidth, keepaspectratio]{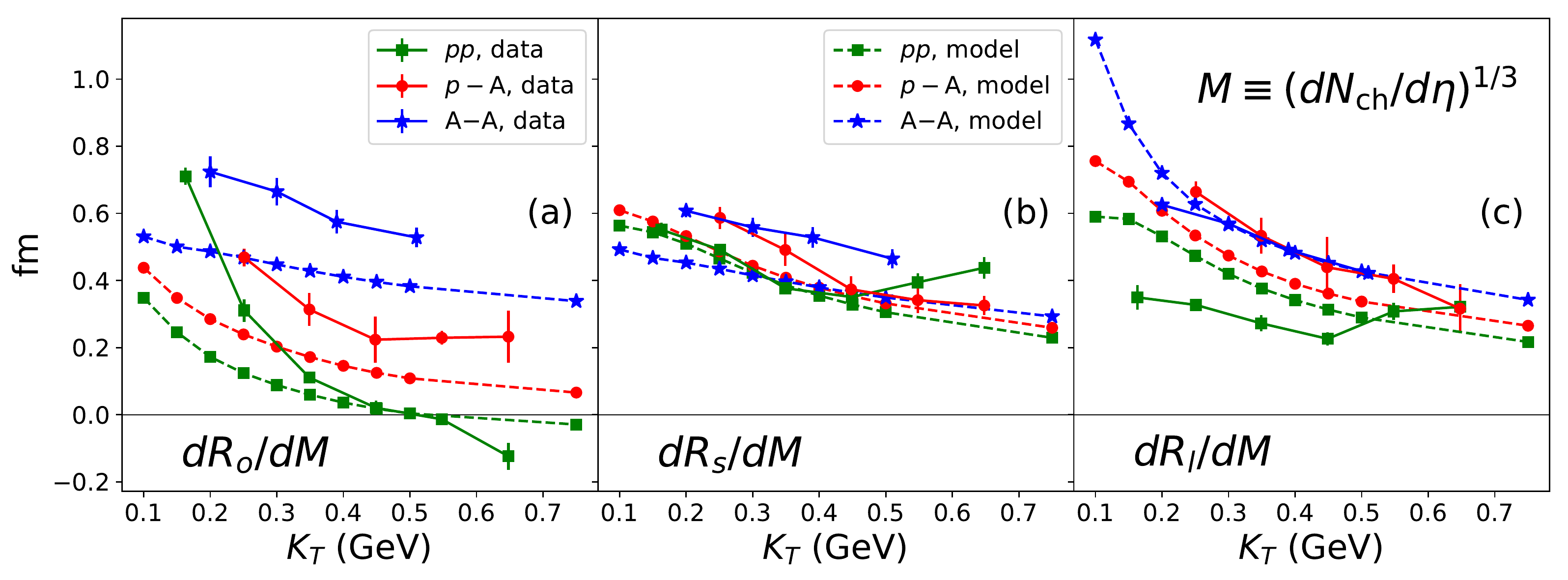}
\caption{The $(dN_\ch/d\eta)^{1/3}$-slopes in the HBT radii extracted from the high-multiplicity model calculations presented here for \pp, \pPb, and \PbPb (dashed lines), compared with corresponding fits to the experimental datasets presented in Fig.~\ref{Fig:WorldData}.  Examples of the fits are given by the dashed lines in Figs.~\ref{Fig:WorldData} and \ref{Fig:R2ij_vs_dNchdeta}b. \label{Fig:model_to_data_slopes}}
\end{figure*}

The result of this (somewhat heuristic) fitting exercise is plotted in Fig.~\ref{Fig:model_to_data_slopes}.  It needs to be reiterated that this should not be taken as a serious, quantitative comparison between the model results computed in this work and the experimental data.  Rather, this is an efficient way of assessing to what extent the high-multiplicity, flow-driven behavior exhibited by small systems is capable, at least in principle, of explaining the non-trivial features observed in the $(dN_\ch/d\eta)^{1/3}$-dependence of the HBT radii.  The slopes $dR_{ij}/dM$, with $M \equiv (dN_\ch/d\eta)^{1/3}$, are shown for each radius and collision system used in this study.  Note that the experimental slopes for \AA have actually been taken from the STAR datasets for Au+Au and Cu+Cu collisions, whereas the model \AA slopes were taken from the \PbPb radii plotted in Fig.~\ref{Fig:R2ij_vs_dNchdeta}.  Since the slopes of the radii are approximately universal in \AA, this approximation is a reasonable one \cite{Lisa:2005dd}.

The features of non-universality and hierarchy in the slopes can now be clearly seen.  Non-universality is manifested by comparing curves in the same panel, which reveals how the slope changes across collision systems.  For instance, panel (a) show that in both the data and the model there is a dramatic splitting of the $R_o$ slope in \pp, \pA, and \AA, with the magnitude of the slopes ordered by system size (and the \pp scaling even turning negative at a certain $K_T$).  By constrast, the slope of $R_s$ is approximately universal in all systems, in both the data and the model.  Perhaps the largest tension between the model and data on this count occurs in $R_l$, for which the slopes in the \pp data are considerably lower than those in \pA and \AA, while the corresponding model calculations show no such dramatic splitting.  Even so, the problem could still be more quantitative than qualitative, and a model which relaxed the assumption of boost invariance would likely see an improvement of the agreement with $R_l$.

What I have here called the `slope hierarchy' is also easily visible in Fig.~\ref{Fig:model_to_data_slopes} by comparing same color curves in different panels, i.e., by comparing the behavior of \textit{different} radii in the \textit{same} system.  Here the verdict is generally encouraging: the $R_o$ slope is generically far smaller than that of $R_s$ or $R_l$, in both model and data.  In general, we also find that the slopes mostly decrease with $K_T$, in both model and data (with a few exceptions).  This is what one should expect to find: larger multiplicities lead to more distorted geometries (cf.~Fig.~\ref{Fig:FOsurfaces}) which tend to exacerbate the associated $K_T$ scaling which probes close to the edge of the system (cf.~Fig.~\ref{Fig:emission_densities}).

Finally, it has to be emphasized yet again that the failure of the non-universality and hierarchy of slopes to persist to small $dN_\ch/d\eta$ is liable to change with the use of a more realistic model.  It is also certainly interesting to consider the possibility that hydrodynamics is valid in \pp only at sufficiently high multiplicities, but merges smoothly to some alternative formulation at smaller multiplicities.  Nevertheless, regardless of whether hydrodynamics turns out to be valid at low multiplicities, it is still the case that the scaling at sufficiently large multiplicities reproduces much of the observed slope hierarchy and non-universality of the radii in a natural and automatic way.  Whether the situation at low multiplicities improves once it is coupled with state-of-the-art hydrodynamic simulations will be investigated in a future study.

\section{Conclusions}
\label{Sec5}

In this paper, I have considered the multiplicity dependence of the HBT radii in detail from the perspective of a simplified, hydrodynamic model.  The advantage of using such a simplified model is that the most important conclusions are easy to draw.  In this case, by comparing the radii measured in large and small systems \textit{at a fixed multiplicity}, one sees that, according to hydrodynamics, the strength of collective flow in the latter systems is disproportionately stronger than in the former systems, leading to measurable effects on the structure of the freeze-out surface itself.

The fact that $R^2_o$ contains a mixture of space and time information, whereas $R^2_s$ and $R^2_l$ are dominated mainly by spatial geometry, implies that the multiplicity scaling in the out direction in small systems should be substantially weaker than in the side or long directions in systems with strong collective flow.  The slope hierarchy observed in the data is therefore an automatic consequence of the hydrodynamic paradigm presented here, suggesting that the evolution of high-multiplicity \pp collisions may be driven by a violent, hydrodynamic response to initial density gradients in these systems.  More generally, it is clear that the multiplicity dependence of HBT can be used to place non-trivial constraints on any model of collectivity in nuclear collisions, and may provide additional discriminating power for adjudicating between the various mechanisms which have been proposed to explain it.  Although a number of features in the calculations outlined above will certainly change quantitatively with additional theoretical improvements, the primary connections between strong collective flow in small systems, the non-trivial evolution of the freeze-out surfaces in large and small systems with multiplicity, and the resulting effects on the HBT radii, are all expected to survive a more rigorous analysis.

Nevertheless, the primary drawback of employing a highly simplified model like the one used here is that one needs to relax a number of strong assumptions and approximations before the model's implications can be taken seriously in a quantitative way.  In this vein, there remain several important directions in which this work will be extended, some of which are already underway.

First, a more sophisticated treatment of initial state fluctuations, viscous corrections, and the inclusion of resonance decay effects and a hadronic rescattering phase are features which must be incorporated before a quantitative comparison with data may be legitimately performed.  Moreover, as pointed out in \cite{Bozek:2013df}, systematic uncertainties arising from the parameterization and construction of the correlation function have not been treated in this work and must in general be handled with great care \cite{Frodermann:2006sp, Plumberg:2016sig}.

Second, one should consider alternatives to hydrodynamics which might also be capable of reproducing the essential behavior seen in the data.  One example is Pythia \cite{Sjostrand:2006za, *Sjostrand:2014zea} and its recent extension, Angantyr \cite{Bierlich:2018xfw}, to the description of collective effects in both small and large collision systems \cite{PlumbergSjostrandLonnblad}.  Alternatively, one could consider to what extent $K_T$ scaling and $(dN_\ch/d\eta)^{1/3}$ scaling of the femtoscopic radii could reflect collectivity which is generated by models with initial-state correlations, such as CGC/IP-Glasma \cite{McLerran:2013xba, Bzdak:2013zma, Schenke:2014zha}.

In the latter case, initial-state models, which predict some flow already at early times, suggest different kinematic initial conditions in nuclear collisions than those which generate the same flow dynamically throughout the collision history, i.e., in response to initial density gradients.  This difference would be undetectable using momentum-space (flow) information alone, but might be identifiable by examining suitable momentum-space observables (e.g., $\l< p_T \r>$, $v_2$) simultaneously with complementary coordinate-space observables (e.g., azimuthally sensitive $R^2_{ij}(K_T, \Phi_K)$).  For this reason, for a fixed amount of flow and multiplicity, initial-state models will generically predict smaller source radii than hydrodynamic models.  Conditioning on both momentum-space and space-time information simultaneously therefore offers constraining power which the use of momentum-space information alone does not.

Third, one should in principle consider the effects of replacing the mid-rapidity charge particle density with forward/backward multiplicity estimators, for which the scaling of the radii could change significantly and might also reveal distinct insights into the systems' dynamics.  Doing so quantitatively would of course require relaxing the assumption of boost invariance used here, and would involve solving the full baryon density equation of motion in conjunction with the usual hydrodynamics in 3+1 dimensions.

Fourth and finally, one can also explore whether additional constraining power can be provided by studying higher moments of the distributions of femtoscopic radii which result from the incorporation of event-by-event fluctuations \cite{Plumberg:2015mxa, *Plumberg:2015aaw, Plumberg:2016sig}.  By combining these together with other collections of observables, such as the wide-ranging set of radial and anisotropic flow measurements, one could thereby place non-trivial constraints simultaneously on both the space-time and momentum-space evolution of a wide range of high-energy nuclear collisions.  A major effort along these lines is deferred to future work.

\acknowledgments

The author is grateful to U.~Heinz, L.~L{\"o}nnblad, T.~Sj{\"o}strand, P.~Christiansen, C.~Bierlich, and all of the participants in the ``Third International {\Thorn}ing on QCD Challenges from $pp$ to $AA$" held in Lund, Sweden, for many stimulating conversations.  The author also acknowledges the use of computing resources from Computing resources from both the Minnesota Supercomputing Institute
(MSI) at the University of Minnesota and the Ohio Supercomputer Center
\cite{OhioSupercomputerCenter1987}.  C.~P.~ is funded by the CLASH project (KAW 2017-0036).

\bibliography{References}{}

\begin{thebibliography}{10}

\bibitem{Nagle:2018nvi}
James~L. Nagle and William~A. Zajc.
\newblock {Small System Collectivity in Relativistic Hadronic and Nuclear
  Collisions}.
\newblock {\em Ann. Rev. Nucl. Part. Sci.}, 68:211--235, 2018.

\bibitem{Adolfsson:2020dhm}
J.~Adolfsson et~al.
\newblock {QCD Challenges from pp to A-A Collisions}.
\newblock 3 2020.

\bibitem{Martinez:2018tuf}
Mauricio Martinez, Matthew~D. Sievert, and Douglas~E. Wertepny.
\newblock {Multiparticle Production at Mid-Rapidity in the Color-Glass
  Condensate}.
\newblock {\em JHEP}, 02:024, 2019.

\bibitem{Wertepny:2020jun}
Douglas Wertepny, Jacquelyn Noronha-Hostler, Matthew Sievert, Skandaprasad Rao,
  and Noah Paladino.
\newblock {Correlations in the Initial Conditions of Heavy-Ion Collisions}.
\newblock {\em EPJ Web Conf.}, 235:08002, 2020.

\bibitem{Lin:2015ucn}
Zi-Wei Lin, Liang He, Terrence Edmonds, Feng Liu, Denes Molnar, and Fuqiang
  Wang.
\newblock {Elliptic Anisotropy $v_2$ May Be Dominated by Particle Escape
  instead of Hydrodynamic Flow}.
\newblock {\em Nucl. Phys. A}, 956:316--319, 2016.

\bibitem{Bierlich:2017vhg}
Christian Bierlich, G{\"o}sta Gustafson, and Leif L{\"o}nnblad.
\newblock {Collectivity without plasma in hadronic collisions}.
\newblock {\em Phys. Lett. B}, 779:58--63, 2018.

\bibitem{Sjostrand:2018xcd}
Torbj{\"o}rn Sj{\"o}strand.
\newblock {Collective Effects: the viewpoint of HEP MC codes}.
\newblock {\em Nucl. Phys. A}, 982:43--49, 2019.

\bibitem{Kurkela:2018ygx}
Aleksi Kurkela, Urs~Achim Wiedemann, and Bin Wu.
\newblock {Nearly isentropic flow at sizeable $\eta/s$}.
\newblock {\em Phys. Lett. B}, 783:274--279, 2018.

\bibitem{Schenke:2010nt}
Bjoern Schenke, Sangyong Jeon, and Charles Gale.
\newblock {(3+1)D hydrodynamic simulation of relativistic heavy-ion
  collisions}.
\newblock {\em Phys. Rev. C}, 82:014903, 2010.

\bibitem{Gale:2012rq}
Charles Gale, Sangyong Jeon, Björn Schenke, Prithwish Tribedy, and Raju
  Venugopalan.
\newblock {Event-by-event anisotropic flow in heavy-ion collisions from
  combined Yang-Mills and viscous fluid dynamics}.
\newblock {\em Phys. Rev. Lett.}, 110(1):012302, 2013.

\bibitem{Gale:2013da}
Charles Gale, Sangyong Jeon, and Bjoern Schenke.
\newblock {Hydrodynamic Modeling of Heavy-Ion Collisions}.
\newblock {\em Int. J. Mod. Phys. A}, 28:1340011, 2013.

\bibitem{Shen:2014vra}
Chun Shen, Zhi Qiu, Huichao Song, Jonah Bernhard, Steffen Bass, and Ulrich
  Heinz.
\newblock {The iEBE-VISHNU code package for relativistic heavy-ion collisions}.
\newblock {\em Comput. Phys. Commun.}, 199:61--85, 2016.

\bibitem{Weller:2017tsr}
Ryan~D. Weller and Paul Romatschke.
\newblock {One fluid to rule them all: viscous hydrodynamic description of
  event-by-event central p+p, p+Pb and Pb+Pb collisions at $\sqrt{s}=5.02$
  TeV}.
\newblock {\em Phys. Lett. B}, 774:351--356, 2017.

\bibitem{Schenke:2019pmk}
Bjoern Schenke, Chun Shen, and Prithwish Tribedy.
\newblock {Hybrid Color Glass Condensate and hydrodynamic description of the
  Relativistic Heavy Ion Collider small system scan}.
\newblock {\em Phys. Lett. B}, 803:135322, 2020.

\bibitem{Lisa:2005dd}
Michael~Annan Lisa, Scott Pratt, Ron Soltz, and Urs Wiedemann.
\newblock {Femtoscopy in relativistic heavy ion collisions}.
\newblock {\em Ann. Rev. Nucl. Part. Sci.}, 55:357--402, 2005.

\bibitem{Makhlin:1987gm}
A.N. Makhlin and Yu.M. Sinyukov.
\newblock {Hydrodynamics of Hadron Matter Under Pion Interferometric
  Microscope}.
\newblock {\em Z. Phys. C}, 39:69, 1988.

\bibitem{Hirono:2014dda}
Yuji Hirono and Edward Shuryak.
\newblock {Femtoscopic signature of strong radial flow in high-multiplicity
  $pp$ collisions}.
\newblock {\em Phys. Rev. C}, 91(5):054915, 2015.

\bibitem{Graef:2012sh}
Gunnar Graef, Marcus Bleicher, and Qingfeng Li.
\newblock {Examination of scaling of Hanbury-Brown--Twiss radii with charged
  particle multiplicity}.
\newblock {\em Phys. Rev. C}, 85:044901, 2012.

\bibitem{Li:2012np}
Qingfeng Li, Gunnar Graef, and Marcus Bleicher.
\newblock {UrQMD calculations of two-pion HBT correlations in p+p and Pb+Pb
  collisions at LHC energies}.
\newblock {\em J. Phys. Conf. Ser.}, 420:012039, 2013.

\bibitem{Heinz:2019dbd}
Ulrich~W. Heinz and J.~Scott Moreland.
\newblock {Hydrodynamic flow in small systems or: ``How the heck is it possible
  that a system emitting only a dozen particles can be described by fluid
  dynamics?''}.
\newblock {\em J. Phys. Conf. Ser.}, 1271(1):012018, 2019.

\bibitem{Kalaydzhyan:2015xba}
Tigran Kalaydzhyan and Edward Shuryak.
\newblock {Collective flow in high-multiplicity proton-proton collisions}.
\newblock {\em Phys. Rev. C}, 91(5):054913, 2015.

\bibitem{Kisiel:2011jg}
A.~Kisiel.
\newblock {Femtoscopy of Pb-Pb and pp collisions at the LHC with the ALICE
  experiment}.
\newblock {\em J. Phys. G}, 38:124008, 2011.

\bibitem{Aamodt:2011kd}
K.~Aamodt et~al.
\newblock {Femtoscopy of $pp$ collisions at $\sqrt{s}=0.9$ and 7 TeV at the LHC
  with two-pion Bose-Einstein correlations}.
\newblock {\em Phys. Rev. D}, 84:112004, 2011.

\bibitem{Adamczyk:2014mxp}
L.~Adamczyk et~al.
\newblock {Beam-energy-dependent two-pion interferometry and the freeze-out
  eccentricity of pions measured in heavy ion collisions at the STAR detector}.
\newblock {\em Phys. Rev. C}, 92(1):014904, 2015.

\bibitem{Adam:2015pya}
J.~Adam et~al.
\newblock {Two-pion femtoscopy in p-Pb collisions at $\sqrt{s_{\rm NN}}=5.02$
  TeV}.
\newblock {\em Phys. Rev. C}, 91:034906, 2015.

\bibitem{Bozek:2013df}
Piotr Bozek and Wojciech Broniowski.
\newblock {Size of the emission source and collectivity in ultra-relativistic
  p-Pb collisions}.
\newblock {\em Phys. Lett. B}, 720:250--253, 2013.

\bibitem{McLerran:2013xba}
Larry McLerran.
\newblock {The Color Glass Condensate, Glasma and the Quark Gluon Plasma in the
  Context of Recent pPb Results from LHC}.
\newblock {\em J. Phys. Conf. Ser.}, 458:012024, 2013.

\bibitem{Wiedemann:1999qn}
Urs~Achim Wiedemann and Ulrich~W. Heinz.
\newblock {Particle interferometry for relativistic heavy ion collisions}.
\newblock {\em Phys. Rept.}, 319:145--230, 1999.

\bibitem{Heinz:1996bs}
Ulrich~W. Heinz.
\newblock {Hanbury-Brown/Twiss interferometry for relativistic heavy ion
  collisions: Theoretical aspects}.
\newblock In {\em {International Summer School on Correlations and Clustering
  Phenomena in Subatomic Physics}}, pages 137--177, 8 1996.

\bibitem{Pratt:1997pw}
S.~Pratt.
\newblock {Validity of the smoothness assumption for calculating two-boson
  correlations in high-energy collisions}.
\newblock {\em Phys. Rev. C}, 56:1095--1098, 1997.

\bibitem{Wiedemann:1996ig}
Urs~Achim Wiedemann and Ulrich~W. Heinz.
\newblock {Resonance contributions to HBT correlation radii}.
\newblock {\em Phys. Rev. C}, 56:3265--3286, 1997.

\bibitem{Sinyukov:1994en}
Yu.M. Sinyukov and Y.Yu. Tolstykh.
\newblock {Coherence influence on the Bose-Einstein correlations}.
\newblock {\em Z. Phys. C}, 61:593--597, 1994.

\bibitem{Sinyukov:2012ut}
Yu.M. Sinyukov and V.M. Shapoval.
\newblock {Correlation femtoscopy of small systems}.
\newblock {\em Phys. Rev. D}, 87(9):094024, 2013.

\bibitem{Akkelin:2011zz}
S.V. Akkelin and Yu.M. Sinyukov.
\newblock {Simple estimates of non-femtoscopic particle correlations in p + p
  collisions}.
\newblock {\em Phys. Part. Nucl. Lett.}, 8(9):959--964, 2011.

\bibitem{Shapoval:2013jca}
V.M. Shapoval, P.~Braun-Munzinger, Iu.~A. Karpenko, and Yu.~M. Sinyukov.
\newblock {Femtoscopic scales in $p+p$ and $p+$Pb collisions in view of the
  uncertainty principle}.
\newblock {\em Phys. Lett. B}, 725:139--147, 2013.

\bibitem{Heinz:1999rw}
Ulrich~W. Heinz and Barbara~V. Jacak.
\newblock {Two particle correlations in relativistic heavy ion collisions}.
\newblock {\em Ann. Rev. Nucl. Part. Sci.}, 49:529--579, 1999.

\bibitem{Heinz:2004qz}
Ulrich~W. Heinz.
\newblock {Concepts of heavy ion physics}.
\newblock In {\em {2nd CERN-CLAF School of High Energy Physics}}, pages
  165--238, 7 2004.

\bibitem{Plumberg:2015eia}
Christopher Plumberg and Ulrich Heinz.
\newblock {Interferometric signatures of the temperature dependence of the
  specific shear viscosity in heavy-ion collisions}.
\newblock {\em Phys. Rev. C}, 91(5):054905, 2015.

\bibitem{Cooper:1974mv}
Fred Cooper and Graham Frye.
\newblock {Comment on the Single Particle Distribution in the Hydrodynamic and
  Statistical Thermodynamic Models of Multiparticle Production}.
\newblock {\em Phys. Rev. D}, 10:186, 1974.

\bibitem{McNelis:2019auj}
Mike McNelis, Derek Everett, and Ulrich Heinz.
\newblock {Particlization in fluid dynamical simulations of heavy-ion
  collisions: The iS3D module}.
\newblock 12 2019.

\bibitem{Paquet:2020rxl}
J.-F. Paquet et~al.
\newblock {Revisiting Bayesian constraints on the transport coefficients of
  QCD}.
\newblock In {\em {28th International Conference on Ultrarelativistic
  Nucleus-Nucleus Collisions}}, 2 2020.

\bibitem{Teaney:2003kp}
Derek Teaney.
\newblock {The Effects of viscosity on spectra, elliptic flow, and HBT radii}.
\newblock {\em Phys. Rev. C}, 68:034913, 2003.

\bibitem{Loizides:2014vua}
C.~Loizides, J.~Nagle, and P.~Steinberg.
\newblock {Improved version of the PHOBOS Glauber Monte Carlo}.
\newblock {\em SoftwareX}, 1-2:13--18, 2015.

\bibitem{Bozek:2019wyr}
Piotr Bo\.zek, Wojciech Broniowski, Maciej Rybczynski, and Grzegorz Stefanek.
\newblock {GLISSANDO 3: GLauber Initial-State Simulation AND mOre..., ver. 3}.
\newblock {\em Comput. Phys. Commun.}, 245:106850, 2019.

\bibitem{Adam:2015gka}
Jaroslav Adam et~al.
\newblock {Charged-particle multiplicities in proton--proton collisions at
  $\sqrt{s} = 0.9$ to 8 TeV}.
\newblock {\em Eur. Phys. J. C}, 77(1):33, 2017.

\bibitem{ALICE:2012xs}
Betty Abelev et~al.
\newblock {Pseudorapidity density of charged particles in $p$ + Pb collisions
  at $\sqrt{s_{NN}}=5.02$ TeV}.
\newblock {\em Phys. Rev. Lett.}, 110(3):032301, 2013.

\bibitem{Aamodt:2010cz}
Kenneth Aamodt et~al.
\newblock {Centrality dependence of the charged-particle multiplicity density
  at mid-rapidity in Pb-Pb collisions at $\sqrt{s_{NN}}=2.76$ TeV}.
\newblock {\em Phys. Rev. Lett.}, 106:032301, 2011.

\bibitem{Abbas:2013bpa}
Ehab Abbas et~al.
\newblock {Centrality dependence of the pseudorapidity density distribution for
  charged particles in Pb-Pb collisions at $\sqrt{s_{\rm NN}}$ = 2.76 TeV}.
\newblock {\em Phys. Lett. B}, 726:610--622, 2013.

\bibitem{Huovinen:2009yb}
Pasi Huovinen and Pter Petreczky.
\newblock {QCD Equation of State and Hadron Resonance Gas}.
\newblock {\em Nucl. Phys. A}, 837:26--53, 2010.

\bibitem{Kurkela:2018vqr}
Aleksi Kurkela, Aleksas Mazeliauskas, Jean-François Paquet, Sören
  Schlichting, and Derek Teaney.
\newblock {Effective kinetic description of event-by-event pre-equilibrium
  dynamics in high-energy heavy-ion collisions}.
\newblock {\em Phys. Rev. C}, 99(3):034910, 2019.

\bibitem{Busza:2018rrf}
Wit Busza, Krishna Rajagopal, and Wilke van~der Schee.
\newblock {Heavy Ion Collisions: The Big Picture, and the Big Questions}.
\newblock {\em Ann. Rev. Nucl. Part. Sci.}, 68:339--376, 2018.

\bibitem{Berges:2020fwq}
Jürgen Berges, Michal~P. Heller, Aleksas Mazeliauskas, and Raju Venugopalan.
\newblock {Thermalization in QCD: theoretical approaches, phenomenological
  applications, and interdisciplinary connections}.
\newblock 5 2020.

\bibitem{Pratt:2008qv}
Scott Pratt.
\newblock {Resolving the HBT Puzzle in Relativistic Heavy Ion Collision}.
\newblock {\em Phys. Rev. Lett.}, 102:232301, 2009.

\bibitem{Ahmad:2016ods}
Saeed Ahmad, Hannu Holopainen, and Pasi Huovinen.
\newblock {Dynamical freeze-out criterion in a hydrodynamical description of Au
  + Au collisions at $\sqrt{s_\mathrm{NN}}=200$ GeV and Pb + Pb collisions at
  $\sqrt{s_\mathrm{NN}}=2760$ GeV}.
\newblock {\em Phys. Rev. C}, 95(5):054911, 2017.

\bibitem{Plumberg:2016sig}
Christopher Plumberg and Ulrich Heinz.
\newblock {Hanbury-Brown--Twiss correlation functions and radii from
  event-by-event hydrodynamics}.
\newblock {\em Phys. Rev. C}, 98(3):034910, 2018.

\bibitem{Frodermann:2006sp}
Evan Frodermann, Ulrich Heinz, and Michael~Annan Lisa.
\newblock {Fitted HBT radii versus space-time variances in flow-dominated
  models}.
\newblock {\em Phys. Rev. C}, 73:044908, 2006.

\bibitem{Plumberg:2015mxa}
Christopher Plumberg and Ulrich Heinz.
\newblock {Probing the properties of event-by-event distributions in
  Hanbury-Brown--Twiss radii}.
\newblock {\em Phys. Rev. C}, 92(4):044906, 2015.
\newblock [Addendum: Phys.Rev.C 92, 049901 (2015)].

\bibitem{Adams:2004yc}
J.~Adams et~al.
\newblock {Pion interferometry in Au+Au collisions at S(NN)**(1/2) = 200-GeV}.
\newblock {\em Phys. Rev. C}, 71:044906, 2005.

\bibitem{Abelev:2009tp}
B.I. Abelev et~al.
\newblock {Pion Interferometry in Au+Au and Cu+Cu Collisions at RHIC}.
\newblock {\em Phys. Rev. C}, 80:024905, 2009.

\bibitem{Aamodt:2011mr}
K.~Aamodt et~al.
\newblock {Two-pion Bose-Einstein correlations in central Pb-Pb collisions at
  $\sqrt{{s}_{NN}} =$ 2.76 TeV}.
\newblock {\em Phys. Lett. B}, 696:328--337, 2011.

\bibitem{Sirunyan:2017ies}
Albert~M Sirunyan et~al.
\newblock {Bose-Einstein correlations in $pp, p\mathrm{Pb}$, and PbPb
  collisions at $\sqrt{{s}_{NN}}=0.9-7$ TeV}.
\newblock {\em Phys. Rev. C}, 97(6):064912, 2018.

\bibitem{Sikler:2017mde}
Ferenc Siklér.
\newblock {Femtoscopy with identified hadrons in pp, pPb, and PbPb collisions
  in CMS}.
\newblock {\em Universe}, 3(4):76, 2017.

\bibitem{Graef:2012za}
Gunnar Graef, Qingfeng Li, and Marcus Bleicher.
\newblock {Formation time dependence of femtoscopic $\pi \pi$ correlations in
  p+p collisions at $\sqrt{s_{NN}}$=7 TeV}.
\newblock {\em J. Phys. G}, 39:065101, 2012.

\bibitem{Adam:2014qja}
Jaroslav Adam et~al.
\newblock {Centrality dependence of particle production in p-Pb collisions at
  $\sqrt{s_{\rm NN} }$= 5.02 TeV}.
\newblock {\em Phys. Rev. C}, 91(6):064905, 2015.

\bibitem{Welsh:2016siu}
Kevin Welsh, Jordan Singer, and Ulrich~W. Heinz.
\newblock {Initial state fluctuations in collisions between light and heavy
  ions}.
\newblock {\em Phys. Rev. C}, 94(2):024919, 2016.

\bibitem{Sievert:2019zjr}
Matthew~D. Sievert and Jacquelyn Noronha-Hostler.
\newblock {CERN Large Hadron Collider system size scan predictions for PbPb,
  XeXe, ArAr, and OO with relativistic hydrodynamics}.
\newblock {\em Phys. Rev. C}, 100(2):024904, 2019.

\bibitem{Borsanyi:2013bia}
Szabocls Borsanyi, Zoltan Fodor, Christian Hoelbling, Sandor~D. Katz, Stefan
  Krieg, and Kalman~K. Szabo.
\newblock {Full result for the QCD equation of state with 2+1 flavors}.
\newblock {\em Phys. Lett. B}, 730:99--104, 2014.

\bibitem{Bazavov:2014pvz}
A.~Bazavov et~al.
\newblock {Equation of state in ( 2+1 )-flavor QCD}.
\newblock {\em Phys. Rev. D}, 90:094503, 2014.

\bibitem{Csorgo:1991ej}
T.~Csorgo and S.~Pratt.
\newblock {Structure of the peak in Bose-Einstein correlations}.
\newblock {\em Conf. Proc. C}, 9106175:75--90, 1991.

\bibitem{Chapman:1994ax}
Scott Chapman, Pierre Scotto, and Ulrich~W. Heinz.
\newblock {Model independent features of the two particle correlation
  function}.
\newblock {\em Acta Phys. Hung. A}, 1:1--31, 1995.

\bibitem{Akkelin:1995gh}
S.V. Akkelin and Yu.M. Sinyukov.
\newblock {The HBT interferometry of expanding sources}.
\newblock {\em Phys. Lett. B}, 356:525--530, 1995.

\bibitem{Hanus:2019fnc}
Patrick Hanus, Aleksas Mazeliauskas, and Klaus Reygers.
\newblock {Entropy production in pp and Pb-Pb collisions at energies available
  at the CERN Large Hadron Collider}.
\newblock {\em Phys. Rev. C}, 100(6):064903, 2019.

\bibitem{Sjostrand:2006za}
Torbj{\"o}rn Sj{\"o}strand, Stephen Mrenna, and Peter~Z. Skands.
\newblock {PYTHIA 6.4 Physics and Manual}.
\newblock {\em JHEP}, 05:026, 2006.

\bibitem{Sjostrand:2014zea}
Torbj{\"o}rn Sj{\"o}strand, Stefan Ask, Jesper~R. Christiansen, Richard Corke,
  Nishita Desai, Philip Ilten, Stephen Mrenna, Stefan Prestel, Christine~O.
  Rasmussen, and Peter~Z. Skands.
\newblock {An introduction to PYTHIA 8.2}.
\newblock {\em Comput. Phys. Commun.}, 191:159--177, 2015.

\bibitem{Bierlich:2018xfw}
Christian Bierlich, G{\"o}sta Gustafson, Leif L{\"o}nnblad, and Harsh Shah.
\newblock {The Angantyr model for Heavy-Ion Collisions in PYTHIA8}.
\newblock {\em JHEP}, 10:134, 2018.

\bibitem{PlumbergSjostrandLonnblad}
Christopher Plumberg, Torbj{\"o}rn Sj{\"o}strand, and Leif L{\"o}nnblad.
\newblock {In progress}.

\bibitem{Bzdak:2013zma}
Adam Bzdak, Bjoern Schenke, Prithwish Tribedy, and Raju Venugopalan.
\newblock {Initial state geometry and the role of hydrodynamics in
  proton-proton, proton-nucleus and deuteron-nucleus collisions}.
\newblock {\em Phys. Rev. C}, 87(6):064906, 2013.

\bibitem{Schenke:2014zha}
Bjoern Schenke and Raju Venugopalan.
\newblock {Eccentric protons? Sensitivity of flow to system size and shape in
  p+p, p+Pb and Pb+Pb collisions}.
\newblock {\em Phys. Rev. Lett.}, 113:102301, 2014.

\bibitem{Plumberg:2015aaw}
Christopher Plumberg and Ulrich Heinz.
\newblock {Observable consequences of event-by-event fluctuations of HBT
  radii}.
\newblock {\em Nucl. Phys. A}, 956:381--384, 2016.

\bibitem{OhioSupercomputerCenter1987}
Ohio~Supercomputer Center.
\newblock Ohio supercomputer center, 1987.

\end{thebibliography}
\bibliographystyle{unsrt}

\end{document}